\documentclass[conference]{IEEEtran}
\IEEEoverridecommandlockouts
\usepackage{cite}
\usepackage{amsmath,amssymb,amsfonts}
\usepackage{algorithmic}
\usepackage{graphicx}
\usepackage{textcomp}
\usepackage{xcolor}
\usepackage[utf8]{inputenc} 
\usepackage[T1]{fontenc}    
\usepackage{hyperref}       
\usepackage{url}            
\usepackage{booktabs}       
\usepackage{amsfonts}       
\usepackage{amsmath}        
\usepackage{nicefrac}       
\usepackage{microtype}      
\usepackage{xcolor}         
\usepackage{amssymb}
\usepackage{amsthm}
\usepackage{subcaption}
\usepackage{graphicx}
\usepackage{comment}
\usepackage{bm}
\usepackage{graphicx}
\usepackage{subcaption}
\usepackage{enumitem}

\theoremstyle{plain}
\newtheorem{theorem}{Theorem}
\def\BibTeX{{\rm B\kern-.05em{\sc i\kern-.025em b}\kern-.08em
    T\kern-.1667em\lower.7ex\hbox{E}\kern-.125emX}}
\begin{document}

\title{AudioMoG: Guiding Audio Generation with Mixture-of-Guidance}

\author{
    \IEEEauthorblockN{Junyou Wang*$^{1, 2, 3}$\thanks{* Equal contribution.}, Zehua Chen*$^{1, 2}$, Binjie Yuan*$^{1,2}$, Kaiwen Zheng$^{1}$, \\Chang Li$^{1,2,3}$, Yuxuan Jiang$^{1,2}$, Jun Zhu$^{\dagger1,2}$\thanks{$\dagger$ Corresponding author. E-mail: dcszj@tsinghua.edu.cn.}}
    $\thanks{
    This work is supported by Fundamental and Interdisciplinary Disciplines Breakthrough Plan of the Ministry of Education of China
    (No. JYB2025XDXM101), and the National Natural Science Foundation of China (62550004, U24A20342, U25B6003, 92570001).}$
    \IEEEauthorblockA{$^1$ Tsinghua University, Beijing, China $^2$ Shengshu AI, Beijing, China}
    \IEEEauthorblockA{$^3$ University of Science and Technology of China, Hefei, China}
}

\maketitle

\begin{abstract}
The design of diffusion-based audio generation systems has been investigated from diverse perspectives, such as data space, network architecture, and conditioning techniques, while most of these innovations require model re-training. In sampling, classifier-free guidance (CFG) has been uniformly adopted to enhance generation quality by strengthening condition alignment. However, CFG often compromises diversity, resulting in suboptimal performance. Although the recent autoguidance (AG) method proposes another direction of guidance that maintains diversity, its direct application in audio generation has so far underperformed CFG. In this work, we introduce \textit{AudioMoG}, an improved sampling method that enhances text-to-audio (T2A) and video-to-audio (V2A) generation quality without requiring extensive training resources. We start with an analysis of both CFG and AG, examining their respective advantages and limitations for guiding diffusion models. Building upon our insights, we introduce a mixture-of-guidance framework that integrates diverse guidance signals with their interaction terms (\textit{e.g.}, the unconditional bad version of the model) to maximize cumulative advantages.
Experiments show that, given the same inference speed, our approach consistently outperforms single guidance in T2A generation across sampling steps, concurrently showing advantages in V2A, text-to-music, and image generation.
Demo samples are available at:~\url{audiomog.github.io}.
\end{abstract}

\begin{IEEEkeywords}
diffusion models, classifier-free guidance, mixture-of-guidance, text-to-audio generation, video-to-audio generation
\end{IEEEkeywords}

\section{Introduction}
\label{sec:intro}

Audio generation conditioned on text and video information, known as text-to-audio (T2A) and video-to-audio (V2A) generation, have witnessed significant advancements in recent studies. 
Typically, these systems generate an audio latent in a small space compressed from the audio waveform or the mel-spectrogram, indicated by text embeddings~\cite{kreuk2022audiogen,liu2023audioldm,liu2024audioldm,huang2023make,evans2024fast} or encoded video representations~\cite{xu2024vtaldm,wang2024v2amapper,du2023condfoleygen,luo2023diff}.
Recent efforts have enhanced cross-modal audio generation quality through various perspectives, such as data augmentation~\cite{huang2023make,huang2023make2}, condition information~\cite{jeong2024rewas,wang2024tiva,liu2023audioldm,li2024quality,evans2024fast}, generative models~\cite{kreuk2022audiogen,liu2023audioldm,liu2024audioldm}, network architecture~\cite{huang2023make2,evans2024fast,hung2024tangoflux}, and compression networks~\cite{liu2023audioldm,evans2024fast,evans2025stable}. 
However, most of these improvements require retraining the model from scratch or with significant overhead.

At the sampling stage, classifier-free guidance (CFG) is the dominant approach for improving generation quality by amplifying the effect of text conditions, while its performance remains unsatisfactory due to 
reduced diversity. Recently, alternative guidance mechanisms such as autoguidance (AG)~\cite{karras2024guiding}, which guides a diffusion model with a weak version of itself, have shown promise to outperform CFG in image~\cite{karras2024guiding,zheng2025direct}, motion~\cite{jeon2025spg}, and video~\cite{STG} generation. However, AG has not proven effective in audio generation settings. For instance,
ETTA~\cite{lee2024etta} investigates AG for T2A generation and finds that although it promotes output diversity, it is sensitive to the choice of weak model and fails to match CFG in terms of objective metrics. The potential and proper usage of guidance in cross-modal audio generation remains underexplored for sampling optimization.

In this work, we first revisit the design of guidance strategies in audio generation, where we analyze the behaviors and limitations of the widely-used CFG and recent AG, respectively. We demonstrate that, CFG enhances synthesis quality through an entangled effect of score correction and condition alignment amplification, which complicates independent control over quality and diversity—particularly as improvements in the unconditional model diminish the correction signal. In contrast, AG employs a weaker conditional model to isolate the score correction effect, enabling quality improvements without compromising diversity, though its effectiveness can be sensitive to the choice of the weak model.

Based on these insights, we present AudioMoG, a mixture-of-guidance sampling framework towards enhancing the generation quality without requiring heavy retraining resources, which fully exploits the complementary advantages of diverse guidance methods.
Specifically, AudioMoG presents two mixture strategies: \textit{Parallel Guidance} and \textit{Hierarchical Guidance}, where diverse guidance methods are integrated \textit{in parallel} or \textit{hierarchically}. The first strategy concurrently leverages the benefits of CFG and AG, while it can suffer from suboptimal results due to the potential conflicts behind the behaviors of different guidance methods. With the second strategy, we enable a more structured integration, progressively harnessing their strengths to yield cumulative advantages.
For example, CFG can first emphasize condition alignment for both the strong and weak models, and then AG further strengthens the generation quality by reducing the score estimation errors between CFG-aligned terms.
In AudioMoG, we observe that the bad version of the model can be trained using the same network architecture as the good one but with fewer iterations, or even taken directly from earlier checkpoints, avoiding the dedicated design~\cite{karras2024guiding} or the sensitivity~\cite{lee2024etta} mentioned in previous works.

Our contributions are summarized as follows.
\begin{itemize}[leftmargin=10pt]
\item We make the first attempt to enhance the guidance method of audio generation and improve the generation quality in a computationally lightweight or even training-free manner.
\item We propose a novel guidance method, AudioMoG, where we present a mixture-of-guidance framework to maximize the cumulative advantages of different guidance methods.
\item Experimental validation on diverse cross-modal audio generation and image generation tasks demonstrates that under the same inference speed, AudioMoG consistently outperforms both CFG and AG. Compared to CFG, we improve FAD from 1.76 to 1.38 in T2A~\cite{evans2024fast}, from 0.73 to 0.68 in V2A, and from 2.36 to 1.92 in text-to-music generation. Compared to AG, we improve FID from 1.60 to 1.47 in image generation~\cite{karras2024analyzing}.
\end{itemize} 

\begin{figure}[t] 
  \centering
  \includegraphics[width=\columnwidth]{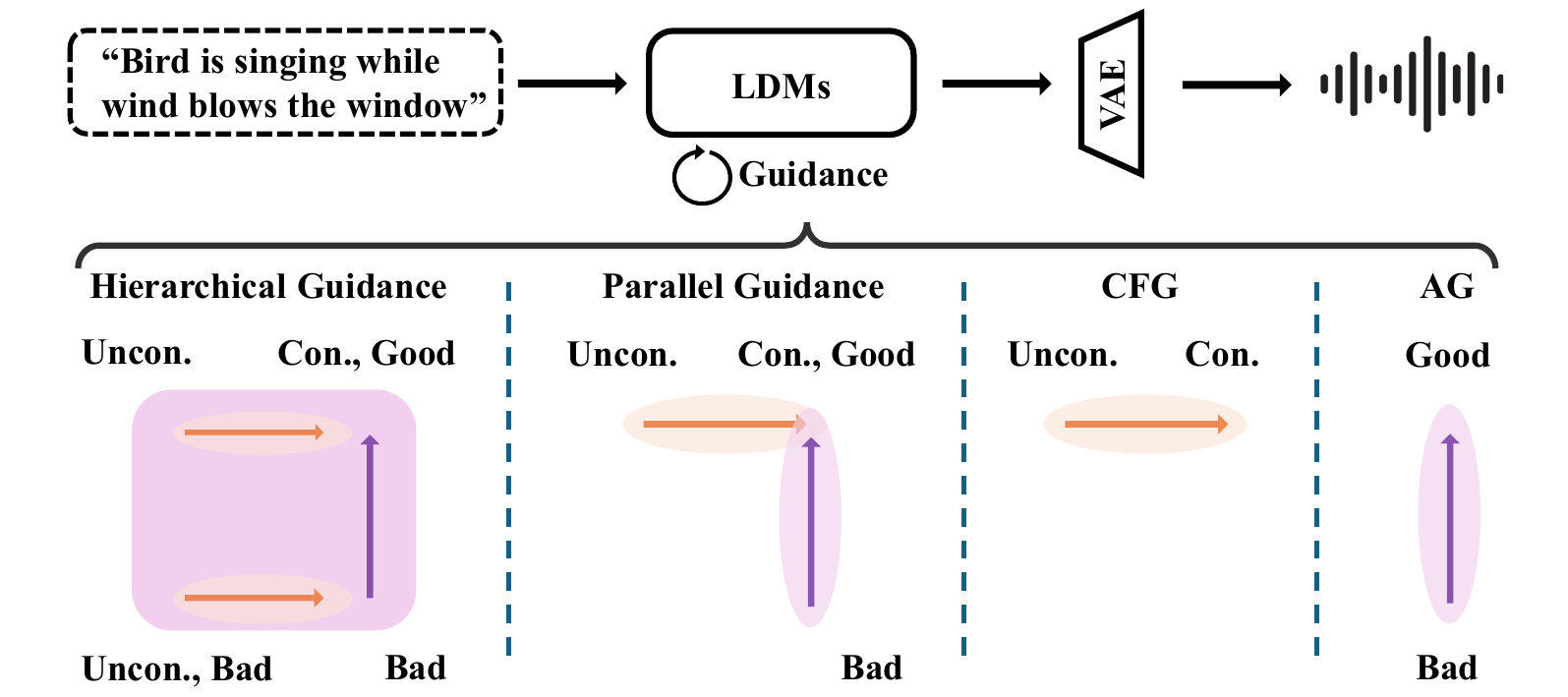}
  \caption{Overall framework of our proposed AudioMoG, which illustrates the mechanism of AudioMoG and its degraded forms—Hierarchical Guidance exploits cumulative advantages from both methods for optimal performance, Parallel Guidance introduces complementary directions, and CFG or AG provides a single-directional guidance.}
  \vspace{-0.2in}
  \label{fig:overview}
\end{figure}

\section{Diffusion-based audio generation system}

In T2A and V2A generation systems, audio signals $\bm{x}$ are first compressed into a small latent space $\bm{z}$ with a compression network. Then, latent diffusion models are popularly adopted to learn the generation of audio latent from a simple prior distribution, \textit{e.g.}, the standard Gaussian noise distribution $\mathcal{N} (\bm{0},\bm{I})$, conditioned on the text prompt or video input.
At the training stage, a forward process is introduced to transform the audio latent at $t=0$ into a noisy latent with
\begin{equation}
    \bm{z}_t = \sqrt{\bar{\alpha}_t} \bm{z}_0 + \sqrt{1-\bar{\alpha}_t} \bm{\epsilon},
\end{equation}
where $\sqrt{\bar{\alpha}_t}$ is predefined to control the signal-to-noise ratio in forward process; $\bm{\epsilon} \sim \mathcal{N}(\bm{0},\bm{I})$ is the added Gaussian noise and shares the same distribution with the prior distribution $p(\bm{z}_T)$ at $t=T$.
At each training iteration, a noise predictor is optimized with
\begin{equation}
    \arg\min_{\theta} \mathbb{E}_{(\bm{z}_0, \bm{c}), \bm{\epsilon}} \left\| \bm{\epsilon} - \bm{\epsilon_\theta} \left( \bm{z}_t, t, \bm{c} \right) \right\|_2^2,
\end{equation}
where $\bm{c}$ is the text embedding or encoded video features to indicate audio generation.
In sampling, each reverse transition $p_{\theta}(\bm{z}_{s}|\bm{z}_{t},\bm{c})$ at the time steps $0 \leq s < t \leq T $ follows a Gaussian distribution $\mathcal{N}(\bm{z}_{s};\mu_{s|t}(\bm{z}_{t},t,\bm{c}),\sigma_{s|t}^{2}\bm{I})$. The mean and variance are parameterized as
\begin{equation}
\label{samplingmeanvar}
\mu_{s|t}(\bm{z}_{t},t,\bm{c})=\sqrt{\bar{\alpha}_{s|t}}(\bm{z}_{t}-\frac{1-\bar{\alpha}_{t|s}}{\sqrt{1-\bar{\alpha}_{t}}}\bm{\epsilon_{\theta}}(\bm{z}_{t},t,\bm{c})),
\end{equation}
\begin{equation}
\sigma^{2}_{s|t}=(1-\bar{\alpha}_{t|s})\frac{1-\bar{\alpha}_{s}}{1-\bar{\alpha}_{t}},
\end{equation}
where $\bar{\alpha}_{t|s}=\bar{\alpha}_{t}/\bar{\alpha}_{s}$.
Given sufficient sampling steps, audio latent $\bm{z}_0$ is reconstructed and then decoded into audio signals $\bm{x}$ with a decoding system.

\section{Guidance methods}
\label{Guidancemethods}

We begin by introducing two representative guidance methods for diffusion models: the widely adopted Classifier-Free Guidance (CFG) and the recently proposed Autoguidance (AG). We then analyze their underlying mechanisms and compare their respective advantages and limitations.

\paragraph{Classifier-Free Guidance}
CFG~\cite{ho2022classifier} is one of the most commonly used strategies in diffusion models for conditional generation. During training, the conditional signal $c$ is randomly replaced with the null condition \( \mathcal{\varnothing} \) with a fixed probability \( p_{\text{uncond}} \) (aka., random label dropout), allowing the model to learn both conditional and unconditional noise predictors, \( \bm {\epsilon_{\theta}}(\bm z_t;t, c) \) and \( \bm {\epsilon_{\theta}}(\bm z_t;t) \). 
At inference time, CFG combines the two predictors as follows:
\begin{equation}
\label{CFG}
    \bm \epsilon_{\text{CFG}}(\bm{z}_t; t, \bm c) = \bm \epsilon_{\theta}(\bm z_t; t) + w \left( \bm \epsilon_{\theta}(\bm z_t; t, \bm c) - \bm \epsilon_{\theta}(\bm z_t; t) \right),
\end{equation}
where \( w \geq 1 \) denotes the guidance scale that adjusts the strength of the conditional signal. 
Larger values of $w$ encourage samples to align more closely with the conditioning signal, potentially enhancing generation quality.

\begin{figure}[t]
  \centering

  \begin{subfigure}[b]{0.24\linewidth}
    \fbox{\includegraphics[width=\linewidth]{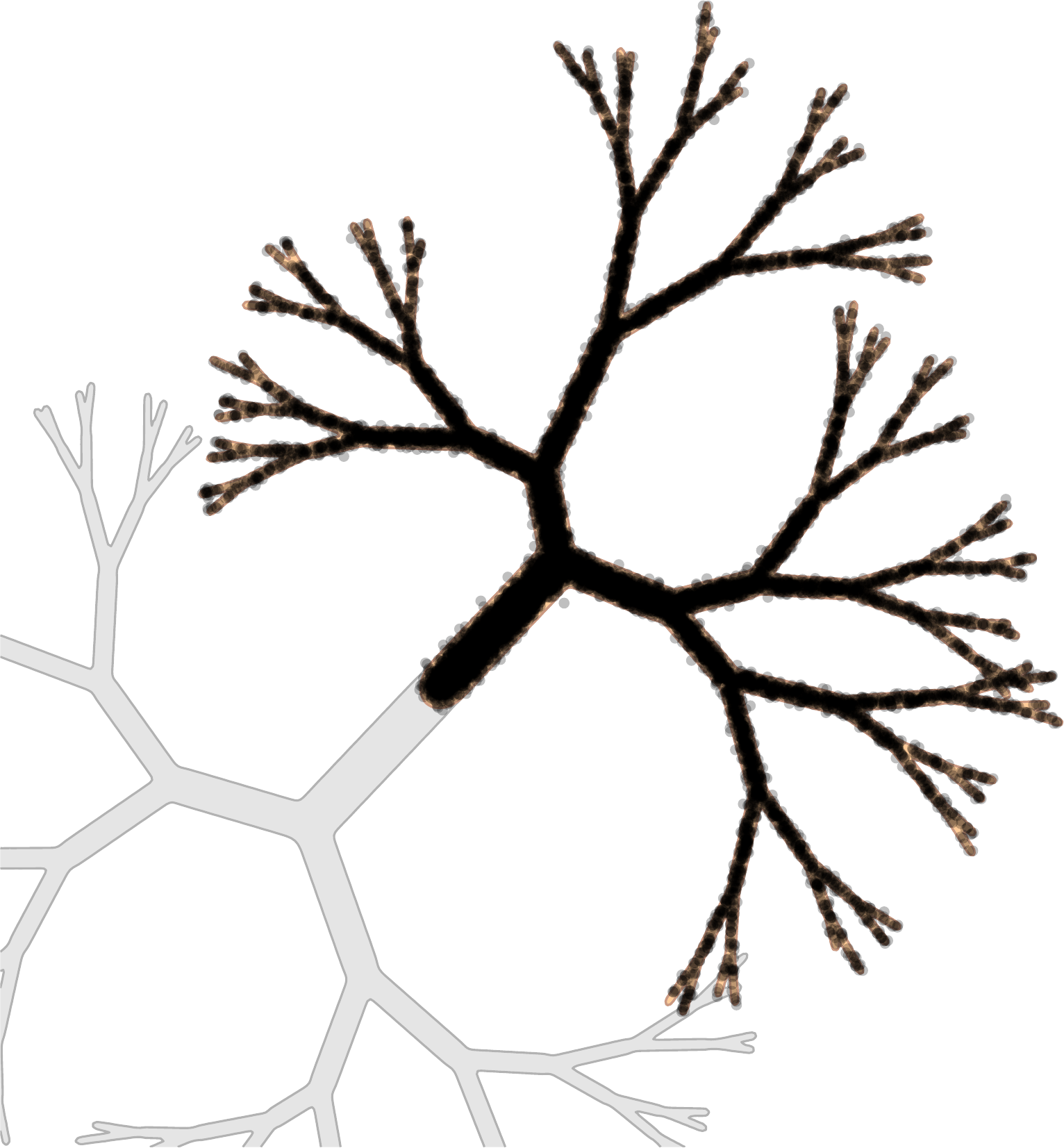}}
    \caption{Ground truth}
    \label{fig:toy:gt}
  \end{subfigure}
  \hspace{4pt}
  \begin{subfigure}[b]{0.24\linewidth}
    \fbox{\includegraphics[width=\linewidth]{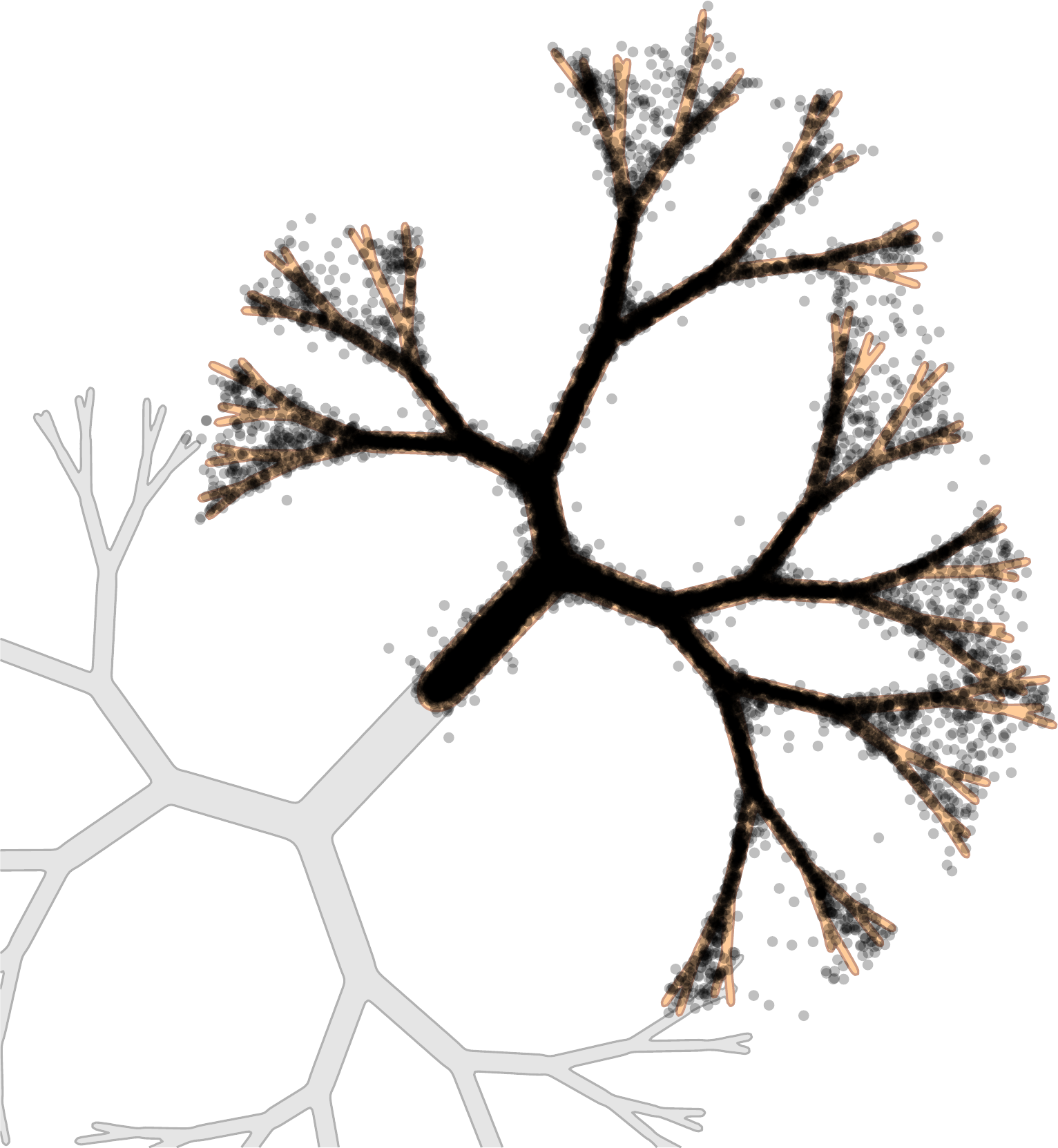}}
    \caption{No guidance}
    \label{fig:toy:ug}
  \end{subfigure}
  \hspace{4pt}
  \begin{subfigure}[b]{0.24\linewidth}
    \fbox{\includegraphics[width=\linewidth]{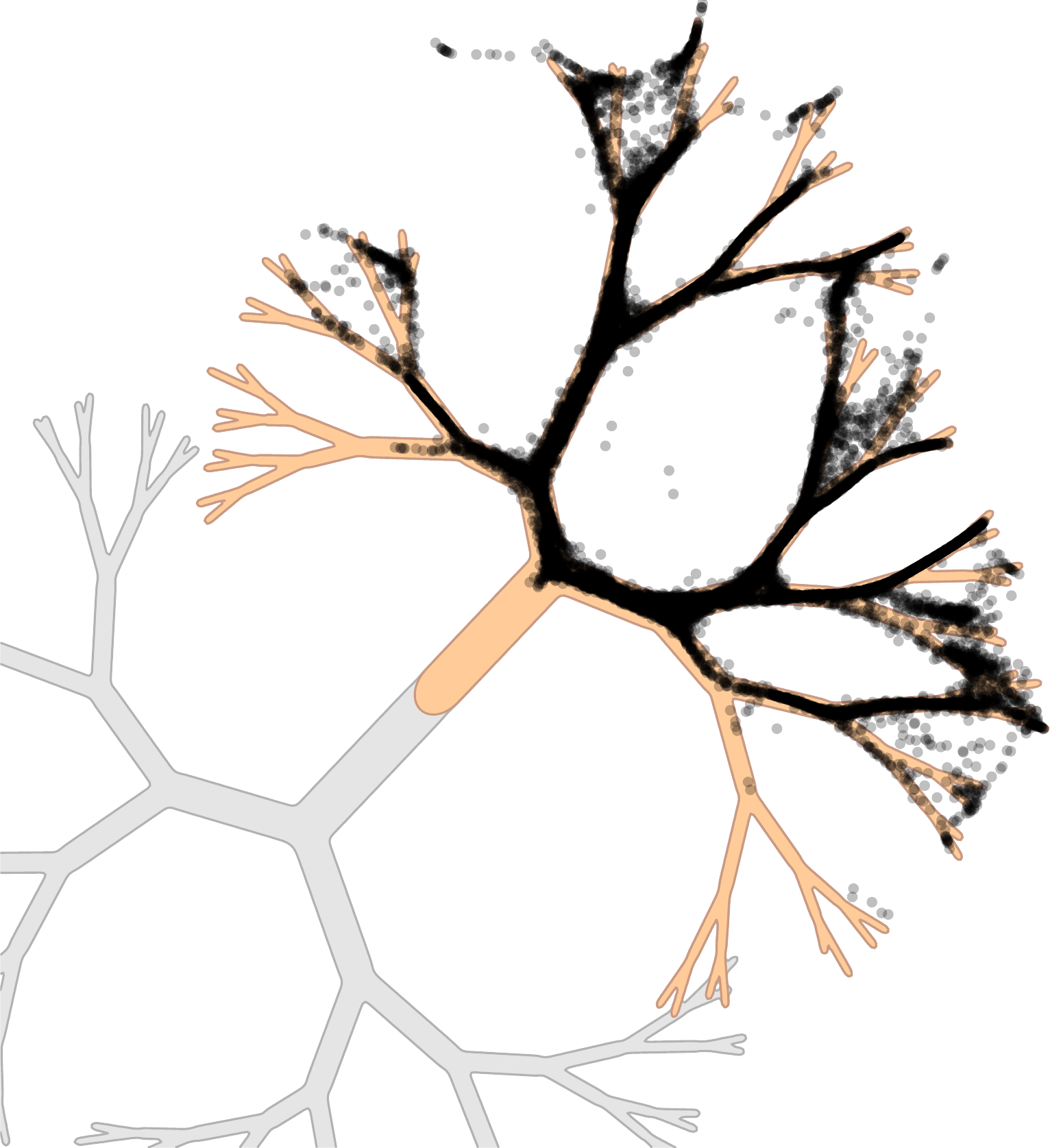}}
    \caption{CFG}
    \label{fig:toy:cfg}
  \end{subfigure}
  \hspace{4pt}
  
  \begin{subfigure}[b]{0.24\linewidth}
    \fbox{\includegraphics[width=\linewidth]{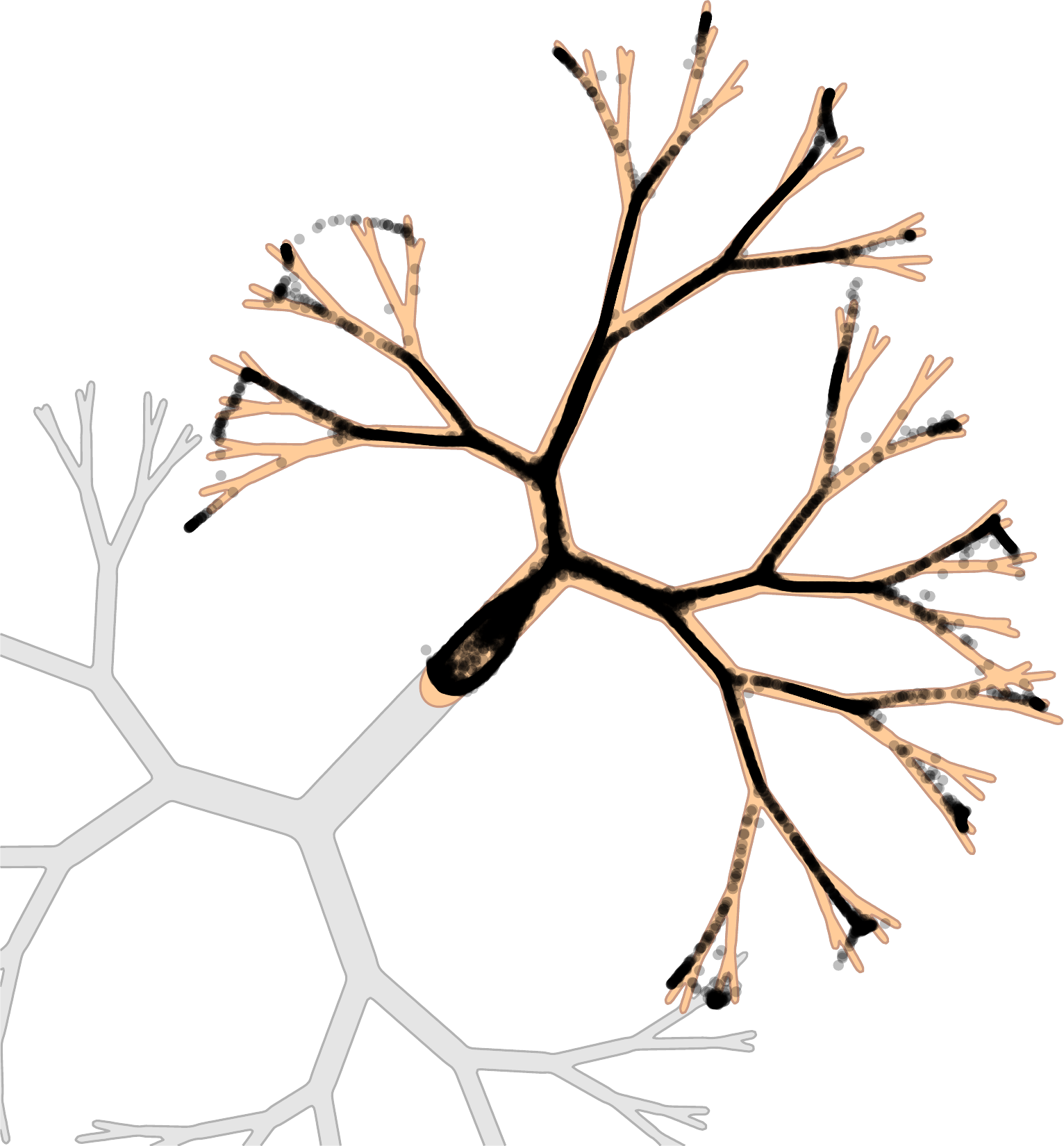}}
    \caption{AG}
    \label{fig:toy:ag}
  \end{subfigure}
  \hspace{4pt}
  \begin{subfigure}[b]{0.24\linewidth}
    \fbox{\includegraphics[width=\linewidth]{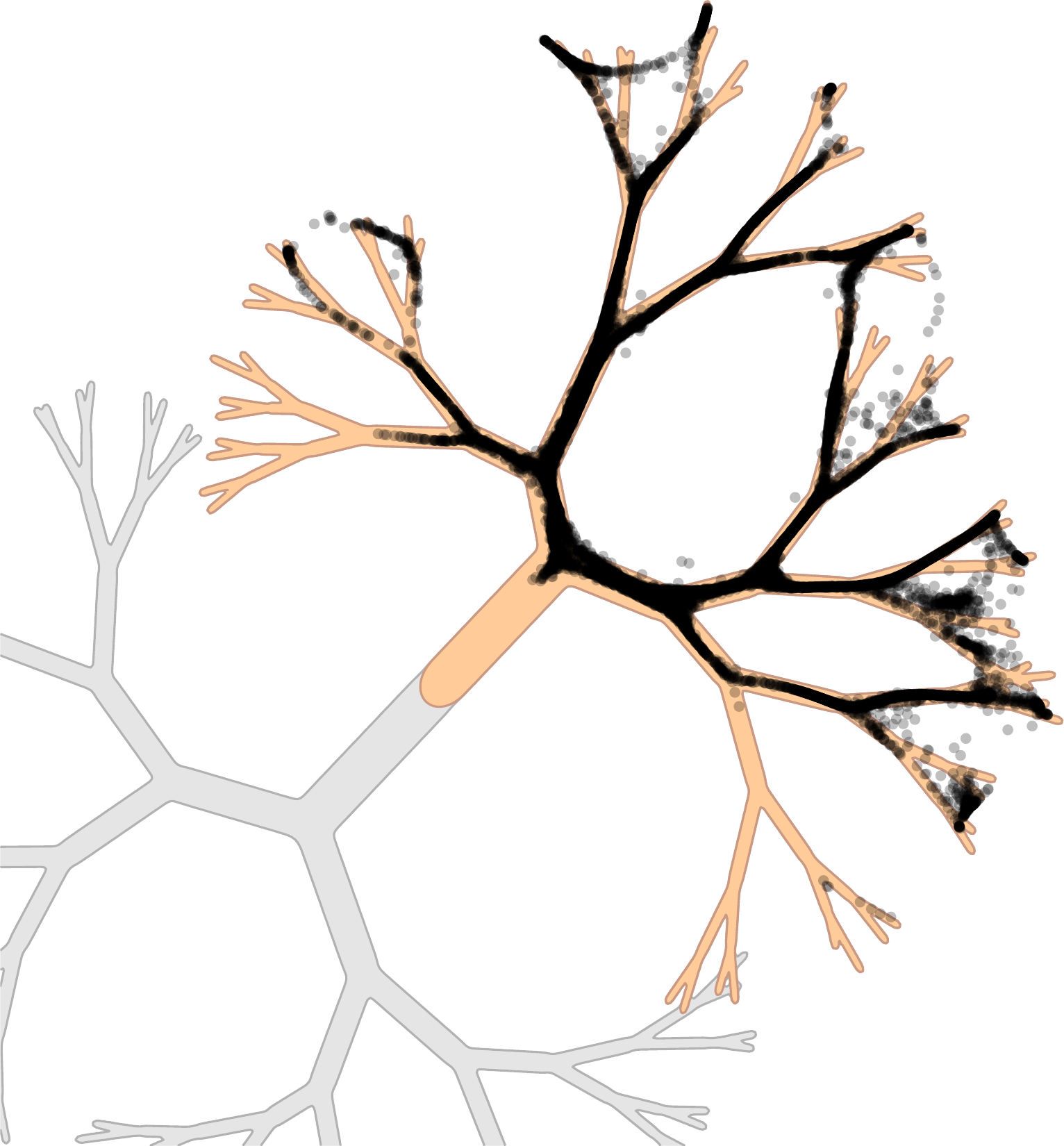}}
    \caption{PG}
    \label{fig:toy:pg}
  \end{subfigure}
  \hspace{4pt}
  \begin{subfigure}[b]{0.24\linewidth}
    \fbox{\includegraphics[width=\linewidth]{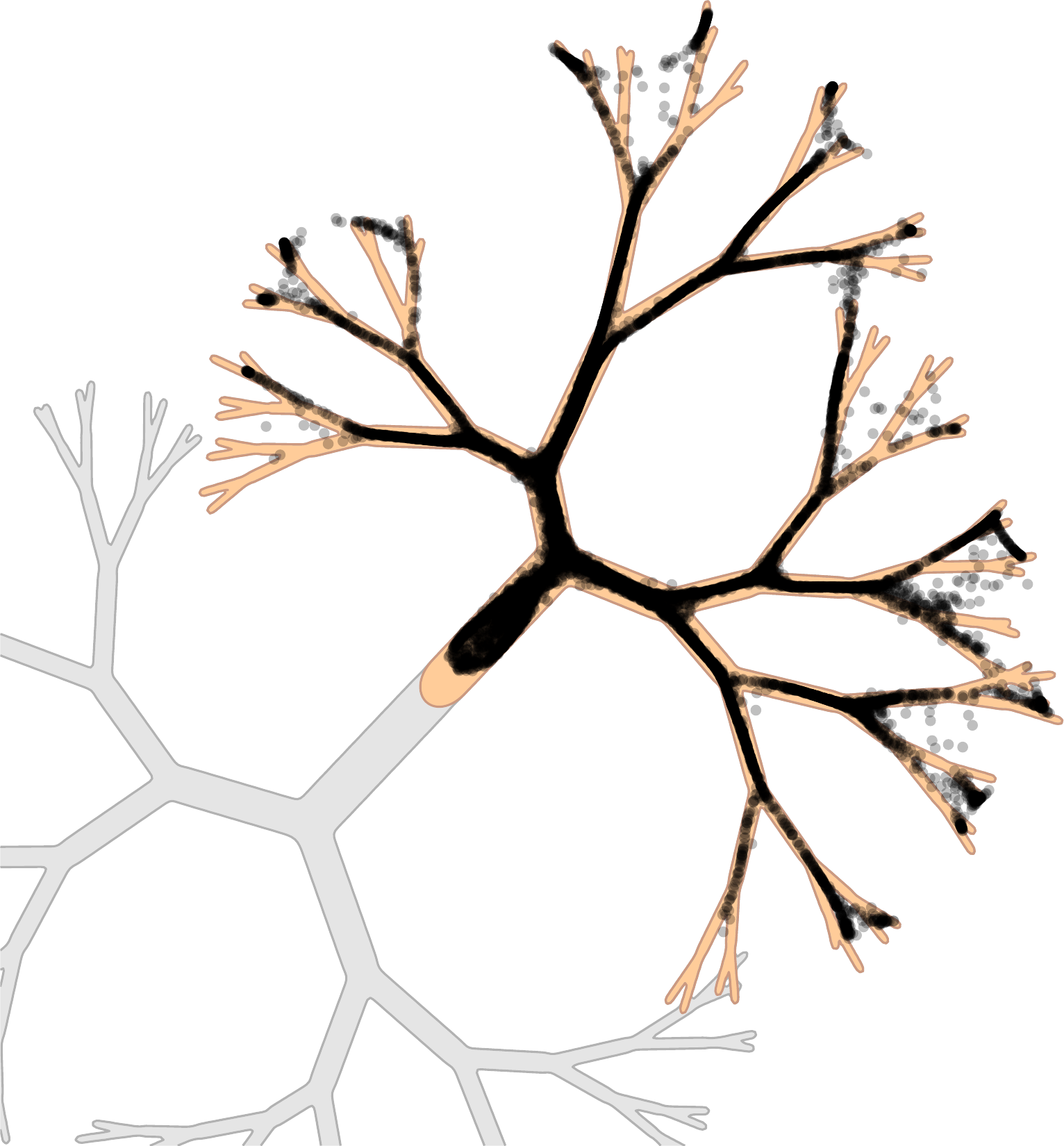}}
    \caption{HG}
    \label{fig:toy:hg}
  \end{subfigure}
  \hspace{4pt}

  \caption{Illustration of guidance methods on the fractal-like 2D distribution from~\cite{karras2024guiding}. 
\textbf{(a)} Ground truth distribution (orange class). 
\textbf{(b)} Unguided conditional sampling generates outliers. 
\textbf{(c)} CFG ($w = 3$) with a well-trained unconditional model struggles to remove outliers. 
\textbf{(d)} AG ($w = 3$) improves score estimation and removes outliers without reducing diversity. 
\textbf{(e)} Parallel Guidance exhibits mode dropping similar to CFG. 
\textbf{(f)} Hierarchical Guidance eliminates outliers and provides more controllable condition alignment.}

  \label{fig:toy}
  \vspace{-0.2in}
\end{figure}

\paragraph{Autoguidance}
AG~\cite{karras2024guiding} proposes guiding a diffusion model using a weaker version of itself:
\begin{equation}
\label{AG}
\bm \epsilon_{\text{AG}}(\bm z_t; t, \bm{c}) = \bm \epsilon_{\theta_{\text{bad}}}(\bm z_t; t, \bm{c}) + w \left( \bm \epsilon_{\theta}(\bm z_t; t, \bm{c}) - \bm \epsilon_{\theta_{\text{bad}}}(\bm z_t; t, \bm{c}) \right),
\end{equation}
where $\theta_{\text{bad}}$ refers to the weak model with smaller size or less training, and $\bm{c}$ can be replaced with $\varnothing$. 
The underlying motivation stems from the observation that the score-matching objective in diffusion models promotes mode coverage, often leading to noisy or inaccurate estimates. By contrastive amplification of the difference between a strong and weak model, AG seeks to improve the score estimation quality. 
Following a similar rationale, recent works~\cite{Kasymov2024AutoLoRAAM, phunyaphibarn2025unconditional, zhong2025domain} leverage the pre-fine-tuned model to guide the model after fine-tuning.
While the specific formulations differ, the core principle remains consistent: leveraging the weak-strong discrepancy to guide improvement.

\paragraph{Analysis}
\label{AnalysisofGuidance}

To demonstrate the mechanisms of CFG and AG, we adopt a 2D toy example introduced in~\cite{karras2024guiding}, where a small denoiser is trained on synthetic data to learn conditional diffusion. Additional details on the experimental setup are provided in Appendix~\ref{2DtoyDetail}. 

The 2D dataset is designed to exhibit low local dimensionality, characterized by highly anisotropic and narrow support, as well as a hierarchical emergence of local detail (Fig.~\ref{fig:toy:gt}), mimicking real-world data manifolds~\cite{karras2024guiding}. Due to limited capacity, the small denoiser network learns a suboptimal score function, leading to scattered and unlikely outliers under unguided sampling (Fig.~\ref{fig:toy:ug}).

CFG improves sample quality by contrasting the conditional and unconditional models. The unconditional model is arguably relatively under-trained due to the inherent difficulty of the unconditional task and the low label dropout rate (usually 10\%). Consequently, the CFG effect in~\eqref{CFG} is mixed, as formulated in the following score decomposition:
\begin{align}
\label{AG2}
\nabla_{\bm x} \log p(\bm x | \bm c)
- \nabla_{\bm x} \log p(\bm x | \varnothing)
=
\bigl[\nabla_{\bm x} \log \mathbb{E}_c\, p(\bm x | \bm c) \nonumber \\
- \nabla_{\bm x} \log p(\bm x | \varnothing)\bigr]  
+ \nabla_{\bm x} \log p(\bm c | \bm x).
\end{align}

With an under-trained unconditional model, the CFG direction entangles two components: (1) \textbf{score correction} from weak-strong contrast that eliminates dispersed outliers, and (2) \textbf{condition alignment amplification} that may skew the distribution and reduce diversity. The quality improvement observed with CFG can be ideally attributed to the first factor. However, the entanglement makes CFG difficult to independently control diversity and quality, as a better unconditional model yields weaker signals for score correction. This effect is visualized in Fig.~\ref{fig:toy:cfg}, where CFG with a sufficiently trained unconditional model fails to eliminate dispersion.

AG, on the other hand, avoids this entanglement by employing a weaker but still conditional version of the model, thus isolating the improvement direction without losing diversity (Fig.~\ref{fig:toy:ag}). 
However, its effectiveness hinges on the availability of a suitably degraded weak model. Constructing such a model in practice can be nontrivial, especially when model degradation does not align well with real score estimation errors and the weak-strong contrast cannot provide meaningful directions.
In such cases, incorporating additional sources of guidance may be necessary to achieve more robust quality gains. For example, when training data quality is suboptimal, CFG often yields sharper and more prompt-consistent generations due to its “lower-temperature” behavior~\cite{bradley2024classifier}, creating a more favorable distribution that can complement AG.

\section{AudioMoG}
\label{AudioMoG}

As discussed above, previous guidance methods, CFG and AG, guide the sampling process from distinct perspectives. CFG enhances consistency with conditional information, while AG removes dispersion by mitigating errors. The effectiveness of CFG has been extensively validated in audio generation across various data types~\cite{liu2023audioldm,evans2025stable}, conditional modalities such as text~\cite{deepanway2023text,huang2023make2} and video~\cite{luo2023diff,xu2024vtaldm}, and network architectures~\cite{liu2023audioldm,evans2025stable,li2024quality,hung2024tangoflux}.
However, CFG can still yield suboptimal results due to its overemphasis on condition information. 
Recent work ETTA~\cite{lee2024etta} explores AG on T2A generation but empirically observes limited synthesis quality and strong sensitivity to the choice of weak model. This suggests that the limitations of CFG remain unresolved in T2A tasks.

Motivated by these challenges, we propose a new guidance framework, making the first attempt to introduce the advantages of AG into audio generation.
Our method, AudioMoG, introduces a mixture strategy to leverage the cumulative benefits of diverse guidance mechanisms, thus improving generation quality without sacrificing diversity. Specifically, AudioMoG allows CFG and AG to be combined either~\textit{in parallel} or~\textit{hierarchically}, forming two variants.

\paragraph{Parallel Guidance}
As analyzed in Section~\ref{Guidancemethods} and shown in Fig.~\ref{fig:toy}, CFG and AG result in complementary behaviors. 
To exploit these synergies, a straightforward mixture strategy is to employ them in parallel at each sampling step, which we call~\textit{Parallel Guidance} (\textit{PG} for short).
Specifically, the conditional predictor $\bm{\epsilon_\theta}(\bm{z}_t,t,\bm{c})$ used in~\eqref{samplingmeanvar} is replaced by:
\begin{align}
\label{PG}
\bm{\epsilon}_{\text{PG}}(\bm{z}_t,t,\bm{c})=\bm{\epsilon_\theta}(\bm{z}_t,t)+w_1(\bm{\epsilon_\theta}(\bm{z}_t,t,\bm{c})-\bm{\epsilon_\theta}(\bm{z}_t,t)) \nonumber \\
+w_2(\bm{\epsilon_\theta}(\bm{z}_t,t,\bm{c})-\bm{\epsilon}_{\bm{\theta}_{\text{bad}}}(\bm{z}_t,t,\bm{c})),
\end{align}
where $\bm{\epsilon}_{\bm{\theta}_{\text{bad}}}(\bm{z}_t,t,\bm{c})$ represents the weak version of conditional model (\textit{e.g.}, under-trained), serving as a reference to correct errors in $\bm{\epsilon_\theta}(\bm{z}_t, t, \bm{c})$. 
This formulation unifies CFG and AG: setting $w_2=0$ recovers CFG (Equation~\eqref{CFG}), and setting $w_1=0$ recovers AG (Equation~\eqref{AG}).

The first guidance term (scaled by $w_1$) reinforces alignment with the condition $\bm{c}$. However, excessive reliance on this term can lead to mode collapse, where generations become overly deterministic. In contrast, the second term (scaled by $w_2$) compares the strong and weak models under the same condition, steering the samples away from dispersed regions without skewing them further for condition compliance. Together, the two terms in PG provide complementary corrective signals.

\paragraph{Hierarchical Guidance}
While PG offers a simple and effective integration of CFG and AG, it implicitly assumes their compatibility at each sampling step. However, as discussed in Section~\ref{Guidancemethods}, these two methods guide the model in potentially interfering directions: emphasizing the condition can lead to mode collapse, while correcting from weak references may introduce semantic drift. To better coordinate these behaviors, we present a more structured design,~\textit{Hierarchical Guidance} (HG for short), which applies guidance in a two-stage manner. Specifically, we first apply CFG to both strong and weak models, and then apply AG to their respective outputs\footnote{HG in CFG-AG or AG-CFG orders yields equivalent guidance family, as proven in Appendix~\ref{proof}}:

{\small
\begin{equation}
\begin{aligned}
\label{HG}
\bm{\epsilon}_{\text{CFG}}(\bm{z}_t,t,\bm{c})=\bm{\epsilon}_{\bm{\theta}}(\bm{z}_t,t)&+w_1(\bm{\epsilon_\theta}(\bm{z}_t,t,\bm{c})-\bm{\epsilon}_{\bm{\theta}}(\bm{z}_t,t)),\\
\bm{\epsilon}_{\text{badCFG}}(\bm{z}_t,t,\bm{c})=\bm{\epsilon}_{\bm{\theta}_{\text{bad}}}(\bm{z}_t,t)&+w_2(\bm{\epsilon}_{\bm{\theta}_{\text{bad}}}(\bm{z}_t,t,\bm{c})-\bm{\epsilon}_{\bm{\theta}_{\text{bad}}}(\bm{z}_t,t)), \\
\bm{\epsilon}_{\text{HG}}(\bm{z}_t,t,\bm{c})=\bm{\epsilon}_{\text{badCFG}}(\bm{z}_t,t,\bm{c})&+w_3(\bm{\epsilon}_{\text{CFG}}(\bm{z}_t,t,\bm{c})
-\bm{\epsilon}_{\text{badCFG}}(\bm{z}_t,t,\bm{c})).
\end{aligned}
\end{equation}
}

Here, a new cross-term $\bm{\epsilon}_{\bm{\theta}_{\text{bad}}}(\bm{z}_t,t)$ (\textit{i.e.}, the unconditional bad version) is incorporated compared to PG. This hierarchical structure promotes a more principled interaction between condition- and model-aware signals. In contrast to PG, which only guides the good model using two distinct directions, HG also refines the bad model, yielding a more balanced corrective process. Moreover, HG generalizes PG. When $w_2 = 1$,~\eqref{HG} reduces exactly to~\eqref{PG}, demonstrating the broader expressiveness and adaptability of HG across different sampling regimes.

A potential concern with HG is increased computational cost per sampling step due to multiple model evaluations, as noted in ETTA~\cite{lee2024etta}. Hence, we conduct a comprehensive comparison in Section~\ref{nfe}, matching inference time between CFG and HG. Results show that HG delivers overall better performance under the same inference budget.

\section{Experiments}
\label{experiments}
\subsection{T2A experiment setup}
\label{setup}

\paragraph{Datasets} We mainly use
AudioSet~\cite{gemmeke2017audio}, FSD50k~\cite{fonseca2021fsd50k} and Clotho v2~\cite{drossos2020clotho}. To maintain consistency, each track in these databases was segmented into 10-second clips and resampled at 16~kHz. The details of these datasets are further introduced in the Appendix \ref{datasets}. 
To compare with prior work, we evaluated our models on the widely used AudioCaps benchmark \cite{kim2019audiocaps}, which consists of about 1K 10-second audio clips. 

\paragraph{Model configurations}
We trained the base model for 1~M iterations with a batch size of 8 per GPU. We used the AdamW optimizer with a learning rate of 5e-5 and a condition drop $p_{\text{uncond}}=0.1$ for CFG. The bad model was trained under the same configuration as the main model, but with only 0.1~M iterations. During inference, we use DPM++ 2M SDE \cite{lu2022dpm}. A more detailed description of the model configurations and compression networks is provided in the Appendix \ref{config}. 

\paragraph{Evaluation metrics}
We conduct a comprehensive evaluation of our models using both objective and subjective evaluations to assess audio generation quality and text-audio alignment. Objective metrics include commonly used Fr\'echet Audio Distance (FAD), Kullback-Leibler (KL) divergence, Inception Score (IS), Fr\'echet Distance (FD), and LAION-CLAP score \cite{wu2023clap}. 
For subjective evaluation, we recruited 20 human raters to score two aspects: (i) overall perceptual quality (OVL), and (ii) semantic relevance to the input text (REL). Both scores are rated on a 1–5 scale. More details are introduced in Appendix \ref{objective} and \ref{subjective}, respectively.

\begin{table}[t]
\centering
  \caption{Objective metrics for T2A generation on AudioCaps test set. The best performance for each metric is highlighted in bold, while the second-best is marked with an underline. 
  }
  \vspace{-0.05in}
  \scriptsize
  \begin{tabular}{l | c c c c c}
\toprule
\textbf{Model} & \textbf{FAD $\downarrow$} & \textbf{KL $\downarrow$} & \textbf{IS $\uparrow$} & \textbf{FD $\downarrow$} & \textbf{CLAP $\uparrow$} \\
\midrule
GT & / & / & / & / & 0.52 \\
\midrule
AudioGen \cite{kreuk2022audiogen} & 3.13 & 2.09 & / & / & / \\
AudioGen-Large \cite{kreuk2022audiogen} &  1.82 & 1.69 & / & / & /\\
Make-An-Audio \cite{huang2023make} &  1.61 & 1.61 & 7.29 & \underline{18.32} & / \\
TANGO-AF\&AC-FT-AC \cite{kong2024improving} &  2.54 & / & 11.04 & \textbf{17.19} & / \\
AudioLDM-Large-Full \cite{liu2023audioldm}  & 1.96 & 1.59 & 8.13 & 23.31 & 0.43  \\
AudioLDM 2 \cite{liu2024audioldm}&  2.09 & 1.79 & 8.14 & 26.44 & 0.50 \\
AudioLDM 2-Large \cite{liu2024audioldm} & 1.89 & 1.54 & 8.55 & 	26.18 & \underline{0.53}\\
Stable Audio Open \cite{evans2025stable} & / & 2.14 & / & / & 0.35 \\
\midrule
CFG-only & 1.76 & \textbf{1.44} & 13.46 & 20.94 & \textbf{0.54}\\
MoG-PG (Ours) & \underline{1.54} & \underline{1.47} & \underline{13.47} & 18.50 & \underline{0.53} \\
MoG-HG (Ours) & \textbf{1.38} & \textbf{1.44} & \textbf{13.58} & 18.87 & \textbf{0.54} \\
\bottomrule
\end{tabular}
\label{tab:objective}
\vspace{-0.05in}
\end{table}

\begin{table}[t]
\scriptsize
  \centering
  \setlength{\tabcolsep}{3pt}
  \caption{Subjective metrics for T2A generation on AudioCaps samples.}
  \vspace{-0.05in}
  \label{tab:subjective}
  \begin{tabular}{l| c c c c c}
    \toprule
    \textbf{Metric} & \textbf{GT} & \textbf{AudioLDM} & \textbf{AudioLDM 2} & \textbf{CFG-only} & \textbf{MoG-HG} \\
    \midrule
    \textbf{OVL} $\uparrow$ & $3.23 \pm 0.58$ & $2.26 \pm 0.53$ & $2.76 \pm 0.47$ & $3.20 \pm 0.51$ & \textbf{3.64 $\pm$ 0.48} \\
    \textbf{REL} $\uparrow$ & $3.43 \pm 0.62$ & $2.34 \pm 0.61$ & $2.90 \pm 0.55$ & $3.40 \pm 0.59$ & \textbf{3.90 $\pm$ 0.54} \\
    \bottomrule
  \end{tabular}
  \vspace{-0.1in}
\end{table}

\begin{table}[t]

\centering
  \caption{Objective metrics for V2A generation on VGGSound test set.}
  \vspace{-0.05in}
  \scriptsize
  \begin{tabular}{l | c c c c c c}
\toprule
\textbf{Model} & \textbf{FAD $\downarrow$} & \textbf{KL $\downarrow$} & \textbf{IS $\uparrow$} & \textbf{FD $\downarrow$} & \textbf{IBS $\uparrow$} & \textbf{AA $\uparrow$}\\
\midrule
GT & / & / & / & / & \underline{32.9} & 83.6 \\
\midrule
IM2WAV \cite{sheffer2023hear} & 6.41 & 2.54 & / & / & 19.0 & 74.3 \\
Diff-Foley \cite{luo2023diff} &  5.79 & 3.12 & 10.8 & 21.90 & 20.4 & \textbf{89.9}\\
FoleyGen \cite{mei2024foleygen} &  1.65 & 2.35 & / & / & 26.1 & 73.8 \\
VTA-LDM \cite{xu2024vtaldm} &  2.01 & 2.37 & 10.4 & 12.80 & 26.2 & 77.0 \\
FoleyCrafter \cite{zhang2024foleycrafter}  & 2.32 & 2.54 & 9.9 & 18.10 & 27.7 & 83.6  \\
V2A-Mapper \cite{wang2024v2amapper}&  0.90 & 2.68 & 12.5 & 8.35 & 22.4 & 78.3 \\
VAB-Encodec \cite{su2024vab} & 2.69 & 2.58 & / & / & / & /\\
VATT w/o text \cite{liu2024VATT} & 2.35 & 2.25 & / & / & / & 82.8 \\
\midrule
CFG-only & 0.73 & 2.28 & \underline{17.1} & 4.48 & 32.8 & 85.8\\
MoG-PG (Ours) & \underline{0.70} & \underline{2.22} & 16.8 & \underline{4.14} & \underline{32.9} & 86.1 \\
MoG-HG (Ours) & \textbf{0.68} & \textbf{2.20} & \textbf{17.2} & \textbf{4.06} & \textbf{33.1} & \underline{86.6} \\ 
\bottomrule
\end{tabular}
\label{tab:v2aobjective}
\vspace{-0.05in}
\end{table}

\begin{table}[t]
\scriptsize
\centering
\setlength{\tabcolsep}{2pt}
\caption{Comparison of different guidance methods.}
\vspace{-0.05in}
\label{tab:mog_all}

\begin{subtable}{0.48\linewidth}
\centering
\caption{T2A results}
\vspace{-0.05in}
\label{tab:5.3.1}
\begin{tabular}{c|cccc}
\toprule
\textbf{Method} & \textbf{FAD $\downarrow$} & \textbf{KL $\downarrow$} & \textbf{IS $\uparrow$} & \textbf{FD $\downarrow$}\\
\midrule
No guidance & 7.31 & 2.45 & 5.86 & 38.42 \\
CFG-only    & 1.96 & 1.91 & 7.47 & \textbf{17.40} \\
AG-only     & 2.30 & 1.91 & 7.41 & 17.57 \\
MoG-PG      & 1.67 & 1.54 & 13.52 & 19.01 \\
MoG-HG      & \textbf{1.38} & \textbf{1.44} & \textbf{13.58} & 18.87 \\
\bottomrule
\end{tabular}
\end{subtable}
\hfill
\begin{subtable}{0.48\linewidth}
\centering
\caption{V2A results}
\vspace{-0.05in}
\label{tab:5.3.2}
\begin{tabular}{c|cccc}
\toprule
\textbf{Method} & \textbf{FAD $\downarrow$} & \textbf{KL $\downarrow$} & \textbf{IS $\uparrow$} & \textbf{FD $\downarrow$}\\
\midrule
No guidance & 1.28 & 2.49 & 10.2 & 8.06 \\
CFG-only    & 0.74 & 2.31 & 15.8 & 5.17 \\
AG-only     & 1.06 & 2.37 & 11.4 & 6.84 \\
MoG-PG      & 0.71 & 2.22 & 16.9 & 4.59 \\
MoG-HG      & \textbf{0.68} & \textbf{2.20} & \textbf{17.2} & \textbf{4.06} \\
\bottomrule
\end{tabular}
\end{subtable}

\vspace{-0.1in}
\end{table}

\subsection{Main results}
\label{mainresults}
\paragraph{T2A generation results} We conduct a comparison study of audio generation quality across GT (\textit{i.e.}, ground-truth audio) and a range of systems. The descriptions of these models are further detailed in the Appendix \ref{baseline}. 
For AudioGen and AudioLDM, we report the metrics as presented in their original papers, and for the rest of the methods, we cite the results from ETTA \cite{lee2024etta}. 
In MoG-PG we set $w_1=4.6, w_2=0.2$, while in MoG-HG, we set $w_1=4.0, w_2=3.3,w_3=1.2$.
To demonstrate the effectiveness of our method, we further evaluate the base model using only CFG. The best results are achieved when the CFG scale is set to $w=7$. 
For a fair comparison, we fix the number of function evaluations (NFE) to 400\footnote{This corresponds to 200 sampling steps for CFG and 100 for HG. For PG, we instead keep the number of sampling steps consistent with HG, yielding an NFE of 300, as we observed that further increasing it leads to slight performance degradation for PG.} for both our method and the CFG-only baseline, \textit{ensuring equal inference cost}. All evaluations are conducted on the AudioCaps test set using standard objective metrics for quantitative comparison. The main results are summarized in Table~\ref{tab:objective}. We have the following conclusions:


\textit{Under equal inference resources, our method uniformly outperforms CFG-only.}
In terms of audio quality, our proposed methods demonstrate significantly improved performance compared to CFG-only under the same inference budget. Specifically, both PG and HG outperform CFG-only across all objective metrics. Our results show that improved quality can be achieved \textit{without} increasing inference cost. 


\textit{HG is better than PG.}
Furthermore, we observe that HG consistently outperforms PG. While both methods significantly surpass the CFG-only baseline, HG achieves better performance than PG across all evaluated metrics, indicating better distributional fidelity and text-audio alignment. We hypothesize these improvements stem from the structured formulation of HG, which avoids the potential conflicts present in PG.

\paragraph{V2A generation results}
To further investigate the potential of AudioMoG, we also fine-tune the T2A model to V2A with CLIP features and validate it on the VGGSound \cite{chen2020vggsound} test set. We  evaluated our models using several metrics to assess audio quality, video-audio semantic alignment and temporal alignment, including FAD, KL, IS, FD, ImageBind Score (IBS) \cite{girdhar2023imagebind} and temporal alignment accuracy (AA) introduced in Diff-Foley \cite{luo2023diff}.  Then we compare our results with GT and a variety of systems. The comparative results are summarized in Table \ref{tab:v2aobjective}. A comprehensive improvement across all metrics, including the critical issue of~\textit{temporal alignment} in V2A, demonstrates that our method uniformly enhances cross-modal audio generation. More details about the V2A experiment are provided in Appendix \ref{appv2a}.

\paragraph{Text-to-music and image generation results}
To further demonstrate the effectiveness of our proposed method, we also perform our approach on text-to-music (T2M) generation with our DiT base model, and conditional image generation on the public EDM~\cite{karras2024analyzing} checkpoint. The results are shown in Appendix~\ref{appt2m} and~\ref{appt2i} respectively, which again verifies the efficacy of MoG across different tasks and modalities.

\paragraph{Diverse samplers} In audio generation tasks, we use DPM++ solver, while in image generation, we use the Heun sampler, demonstrating the robustness of our methods on different samplers.


\subsection{Additional results}
\label{nfe}
\begin{figure}[t]
  \centering
  \begin{subfigure}{0.48\linewidth}
    \centering
    \includegraphics[width=\linewidth]{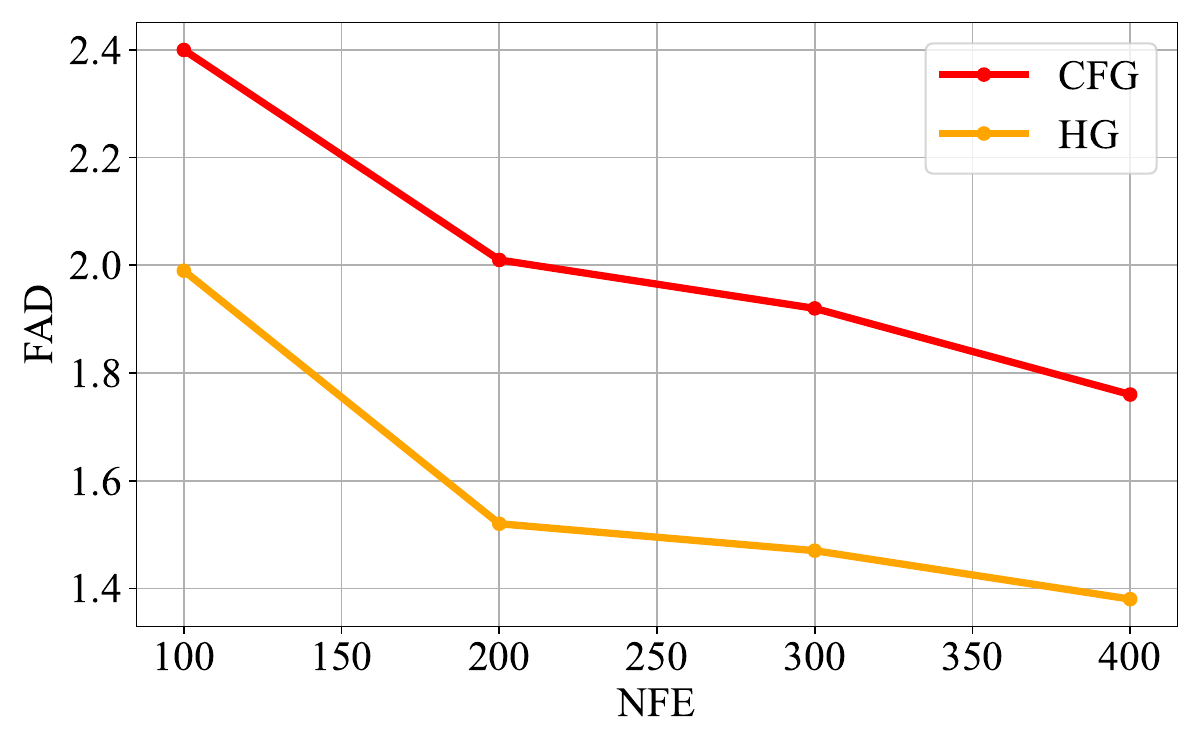}
    \caption{FAD ($\downarrow$) under different NFEs.}
    \label{fig:left}
  \end{subfigure}
  \hfill
  \begin{subfigure}{0.48\linewidth}
    \centering
    \includegraphics[width=\linewidth]{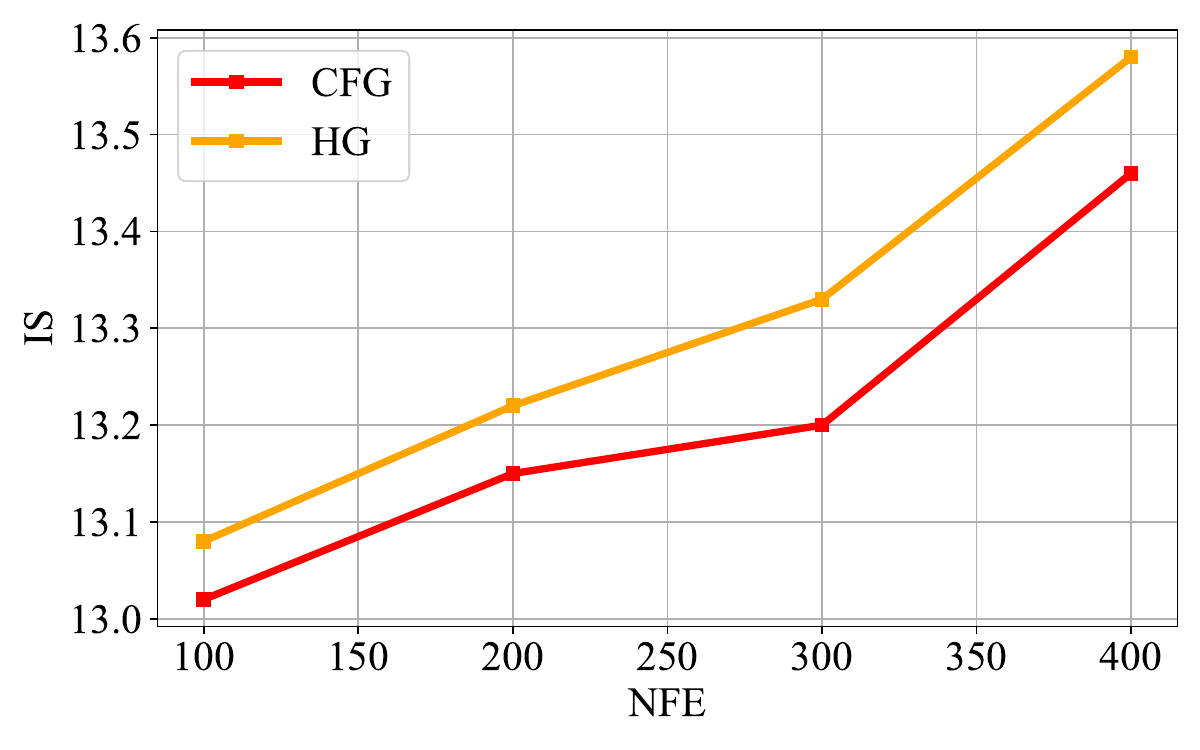}
    \caption{IS ($\uparrow$) under different NFEs.}
    \label{fig:right}
  \end{subfigure}

  \caption{Performance comparison of HG and CFG-only under different NFEs.}
  \label{fig:variousnfes}
  \vspace{-0.2in}
\end{figure}

\paragraph{Comparison across various NFEs}
We compare the performance of HG and CFG-only under different numbers of function evaluations (NFE), each using their respective optimal settings. Specifically, we examine NFE values of 100, 200, 300 and 400, and report the results in Fig.~\ref{fig:variousnfes}. Across all NFE levels, HG consistently outperforms CFG-only in terms of FAD and IS. This consistent superiority indicates that our method is more robust across varying computational budgets, maintaining high-quality generation even under stricter inference constraints. These results highlight the strong adaptability and scalability of HG, which delivers reliable quality improvements regardless of available inference resources, while also unlocking higher performance ceilings when additional computation is allowed. 

\paragraph{Comparison in the same guidance scale}
To validate the effectiveness of our approach, we compare different guidance methods on both T2A and V2A. For NFE, we fix it to 300 for PG and 400 for the rest. For the guidance scale, we set $w_1=w_3=1$, $w_3=1$, $w_1=w_2=1$ and $w_2=1$ in the HG setting, which corresponds to no guidance, CFG-only, AG-only and PG, respectively. The remaining guidance scales are consistent with HG in Table~\ref{tab:objective} and Table~\ref{tab:v2aobjective} (AG-only also achieves the best performance at this scale). The results are shown in Table~\ref{tab:5.3.1} and Table~\ref{tab:5.3.2}. The superiority of PG and HG indicates that combining guidance directions yields better performance than using either one alone. Furthermore, HG surpasses PG on both tasks, further confirming the advantage of the hierarchical structure.

\paragraph{Impact of guidance scales}
We further investigate the influence of guidance scales in PG and HG on the audio generation quality. 
Our analysis reveals that these parameters have an effect on the balance between fidelity and diversity in the generated audio. 
Detailed experimental results and discussions are provided in Appendix \ref{scales}.

\paragraph{Case study}
Apart from the objective and subjective evaluations, we conduct a case study for each of the guidance methods in Appendix~\ref{appgeneratedsamples}. As shown, CFG is prone to produce less structured results, while PG has shown improved quality and HG produces the strongest outcome. 
These results are consistent with our objective test results and the human evaluations.

\section{Conclusion}
In this work, we introduce \textit{AudioMoG}, a mixture-of-guidance framework to improve audio generation quality in the sampling process. In particular, we present two mixture strategies, PG and HG, which directly combine complementary guidance methods, or hierarchically integrate them for cumulative improvements.
Comprehensive experiments demonstrate the superiority of AudioMoG over the popularly adopted CFG on T2A generation under the same inference budgets, as well as achieving improvement over both CFG and AG across V2A, T2M, and image generation tasks without increasing inference time.

{\tiny
\bibliographystyle{IEEEbib}
\bibliography{icme2026references}

@String{Computing = "Computing" }

@ArtifactSoftware{R,
    title = {R: A Language and Environment for Statistical Computing},
    author = {{R Core Team}},
    organization = {R Foundation for Statistical Computing},
    address = {Vienna, Austria},
    year = {2019},
    url = {https://www.R-project.org/},
}

@article{liu2023audioldm,
  title={AudioLDM: Text-to-Audio Generation with Latent Diffusion Models},
  author={Liu et al., Haohe},
  journal={arXiv preprint arXiv:2301.12503},

}

@article{li2024quality,
  title={Quality-aware masked diffusion transformer for enhanced music generation},
  author={Li et al., Chang},
  journal={arXiv preprint arXiv:2405.15863},
}

@article{karras2024guiding,
  title={Guiding a diffusion model with a bad version of itself},
  author={Karras et al., Tero},
  journal={NeurIPS},
  volume={37},
  pages={52996--53021},
  year={2024}
}

@inproceedings{kim2019audiocaps,
  title={Audiocaps: Generating captions for audios in the wild},
  author={Kim et al., Chris Dongjoo},
  booktitle={NAACL-HLT},
  pages={119--132},
  year={2019}
}

@inproceedings{gemmeke2017audio,
  title={Audio set: An ontology and human-labeled dataset for audio events},
  author={Gemmeke et al., Jort F},
  booktitle={ICASSP},
  pages={776--780},
  year={2017},
  organization={IEEE}
}

@inproceedings{drossos2020clotho,
  title={Clotho: An audio captioning dataset},
  author={Drossos, Konstantinos and Lipping, Samuel and Virtanen, Tuomas},
  booktitle={ICASSP},
  pages={736--740},
  year={2020},
  organization={IEEE}
}

@inproceedings{chen2020vggsound,
  title={Vggsound: A large-scale audio-visual dataset},
  author={Chen et al., Honglie},
  booktitle={ICASSP},
  pages={721--725},
  year={2020},
  organization={IEEE}
}

@article{fonseca2021fsd50k,
  title={Fsd50k: an open dataset of human-labeled sound events},
  author={Fonseca et al., Eduardo},
  journal={TASLP},
  volume={30},
  pages={829--852},
  year={2021},
  publisher={IEEE}
}

@article{defferrard2016fma,
  title={FMA: A dataset for music analysis},
  author={Defferrard et al., Micha{\"e}l},
  journal={arXiv preprint arXiv:1612.01840},
}

@inproceedings{bertin2011million,
  title={The million song dataset.},
  author={Bertin-Mahieux et al., Thierry},
  booktitle={Ismir},
  volume={2},
  number={9},
  pages={10},
  year={2011}
}

@inproceedings{law2009evaluation,
  title={Evaluation of algorithms using games: The case of music tagging.},
  author={Law et al., Edith},
  booktitle={ISMIR},
  pages={387--392},
  year={2009},
  organization={Citeseer}
}

@inproceedings{evans2024fast,
  title={Fast timing-conditioned latent audio diffusion},
  author={Evans et al., Zach},
  booktitle={ICML},
  year={2024}
}

@inproceedings{STG,
  title={Spatiotemporal Skip Guidance for Enhanced Video Diffusion Sampling},
  author={Hyung et al., Junha},
  booktitle={CVPR},
  year={2025}
}

@article{liu2024audioldm,
  title={Audioldm 2: Learning holistic audio generation with self-supervised pretraining},
  author={Liu et al., Haohe},
  journal={TASLP},
  year={2024},
  publisher={IEEE}
}

@article{deepanway2023text,
  title={Text-to-audio generation using instruction-tuned llm and latent diffusion model},
  author={Ghosal et al., Deepanway},
  journal={arXiv preprint arXiv:2304.13731},
}

@inproceedings{majumder2024tango,
  title={Tango 2: Aligning diffusion-based text-to-audio generations through direct preference optimization},
  author={Majumder et al., Navonil},
  booktitle={ACMMM},
  pages={564--572},
  year={2024}
}

@inproceedings{huang2023make,
  title={Make-an-audio: Text-to-audio generation with prompt-enhanced diffusion models},
  author={Huang et al., Rongjie},
  booktitle={ICML},
  pages={13916--13932},
  year={2023},
  organization={PMLR}
}

@article{huang2023make2,
  title={Make-an-audio 2: Temporal-enhanced text-to-audio generation},
  author={Huang et al., Jiawei},
  journal={arXiv preprint arXiv:2305.18474},
}

@article{ho2022classifier,
  title={Classifier-free diffusion guidance},
  author={Ho, Jonathan and Salimans, Tim},
  journal={arXiv preprint arXiv:2207.12598},
}

@inproceedings{lee2024etta,
  title={ETTA: Elucidating the Design Space of Text-to-Audio Models},
  author={Lee et al., Sang-gil},
  booktitle={ICML},
  year={2025}
}

@article{xu2024vtaldm,
  title={Video-to-Audio Generation with Hidden Alignment},
  author={Xu et al., Manjie},
  journal={arXiv preprint arXiv:2407.07464},
}

@inproceedings{wang2024tiva,
  title={TiVA: Time-Aligned Video-to-Audio Generation},
  author={Wang et al., Xihua},
  booktitle={ACMMM},
  year={2024}
}

@article{zhang2024foleycrafter,
  title={FoleyCrafter: Bring Silent Videos to Life with Lifelike and Synchronized Sounds},
  author={Zhang et al., Yiming},
  journal={arXiv preprint arXiv:2407.01494},
}

@inproceedings{xie2024sonicvisionlm,
  title={Sonicvisionlm: Playing sound with vision language models},
  author={Xie et al., Zhifeng},
  booktitle={CVPR},
  pages={26866--26875},
  year={2024}
}

@inproceedings{liu2024VATT,
  title={Tell What You Hear From What You See - Video to Audio Generation Through Text},
  author={Liu et al., Xiulong},
  booktitle={NeurIPS},
  year={2024}
}

@inproceedings{wu2023clap,
  title={Large-scale contrastive language-audio pretraining with feature fusion and keyword-to-caption augmentation},
  author={Wu et al., Yusong},
  booktitle={ICASSP},
  pages={1--5},
  year={2023},
  organization={IEEE}
}

@inproceedings{du2023condfoleygen,
  title={Conditional Generation of Audio from Video via Foley Analogies},
  author={Du et al., Yuexi},
  booktitle={CVPR},
  year={2023},
}

@article{jeong2024rewas,
  title={Read, Watch and Scream! Sound Generation from Text and Video},
  author={Jeong et al., Yujin},
  journal={arXiv preprint arXiv:2407.05551},
}

@inproceedings{comunita2024syncfusion,
  title={Syncfusion: Multimodal Onset-Synchronized Video-to-Audio Foley Synthesis},
  author={Comunit{\`a} et al., Marco},
  booktitle={ICASSP},
  pages={936--940},
  year={2024},
  organization={IEEE}
}

@inproceedings{wang2024v2amapper,
  title={V2a-mapper: A lightweight solution for vision-to-audio generation by connecting foundation models},
  author={Wang et al., Heng},
  booktitle={AAAI},
  volume={38},
  number={14},
  pages={15492--15501},
  year={2024}
}

@article{su2024vab,
  title={From vision to audio and beyond: A unified model for audio-visual representation and generation},
  author={Su, Kun and Liu, Xiulong and Shlizerman, Eli},
  journal={arXiv preprint arXiv:2409.19132},
}

@article{yang2023diffsound,
  title={Diffsound: Discrete diffusion model for text-to-sound generation},
  author={Yang et al., Dongchao},
  journal={TASLP},
  volume={31},
  pages={1720--1733},
  year={2023},
  publisher={IEEE}
}

@article{kreuk2022audiogen,
  title={Audiogen: Textually guided audio generation},
  author={Kreuk et al., Felix},
  journal={arXiv preprint arXiv:2209.15352},
}

@article{hung2024tangoflux,
  title={TangoFlux: Super Fast and Faithful Text to Audio Generation with Flow Matching and Clap-Ranked Preference Optimization},
  author={Hung et al., Chia-Yu},
  journal={arXiv preprint arXiv:2412.21037},
}

@article{phunyaphibarn2025unconditional,
  title={Unconditional Priors Matter! Improving Conditional Generation of Fine-Tuned Diffusion Models},
  author={Phunyaphibarn, Prin and Lee, Phillip Y. and Kim, Jaihoon and Sung, Minhyuk},
  journal={arXiv preprint arXiv:2503.20240},
}

@article{jeon2025spg,
  title={SPG: Improving Motion Diffusion by Smooth Perturbation Guidance},
  author={Jeon, Boseong},
  journal={arXiv preprint arXiv:2503.02577},
}

@article{lu2022dpm,
  title={Dpm-solver++: Fast solver for guided sampling of diffusion probabilistic models},
  author={Lu et al., Cheng},
  journal={arXiv preprint arXiv:2211.01095},
}

@article{kong2024improving,
  title={Improving text-to-audio models with synthetic captions},
  author={Kong et al., Zhifeng},
  journal={arXiv preprint arXiv:2406.15487},
}

@inproceedings{evans2025stable,
  title={Stable audio open},
  author={Evans et al., Zach},
  booktitle={ICASSP},
  pages={1--5},
  year={2025},
  organization={IEEE}
}

@article{bradley2024classifier,
  title={Classifier-free guidance is a predictor-corrector},
  author={Bradley, Arwen and Nakkiran, Preetum},
  journal={arXiv preprint arXiv:2408.09000},
}

@article{chidambaram2024does,
  title={What does guidance do? a fine-grained analysis in a simple setting},
  author={Chidambaram et al., Muthu},
  journal={arXiv preprint arXiv:2409.13074},
}

@article{Kasymov2024AutoLoRAAM,
  title={AutoLoRA: AutoGuidance Meets Low-Rank Adaptation for Diffusion Models},
  author={Kasymov et al., Artur},
  journal={ArXiv},
  year={2024},
  volume={abs/2410.03941},
  url={https://api.semanticscholar.org/CorpusID:273186941}
}

@article{zhong2025domain,
  title={Domain guidance: A simple transfer approach for a pre-trained diffusion model},
  author={Zhong et al., Jincheng},
  journal={arXiv preprint arXiv:2504.01521},
}

@article{zheng2025direct,
  title={Direct Discriminative Optimization: Your Likelihood-Based Visual Generative Model is Secretly a GAN Discriminator},
  author={Zheng et al., Kaiwen},
  journal={arXiv preprint arXiv:2503.01103},
}

@inproceedings{sheffer2023hear,
  title={I hear your true colors: Image guided audio generation},
  author={Sheffer, Roy and Adi, Yossi},
  booktitle={ICASSP},
  pages={1--5},
  year={2023},
  organization={IEEE}
}

@article{luo2023diff,
  title={Diff-foley: Synchronized video-to-audio synthesis with latent diffusion models},
  author={Luo et al., Simian},
  journal={NeurIPS},
  volume={36},
  pages={48855--48876},
  year={2023}
}

@inproceedings{mei2024foleygen,
  title={Foleygen: Visually-guided audio generation},
  author={Mei et al., Xinhao},
  booktitle={MLSP},
  pages={1--6},
  year={2024},
  organization={IEEE}
}

@inproceedings{girdhar2023imagebind,
  title={Imagebind: One embedding space to bind them all},
  author={Girdhar et al., Rohit},
  booktitle={CVPR},
  pages={15180--15190},
  year={2023}
}

@article{forsgren2022riffusion,
  title={Riffusion-Stable diffusion for real-time music generation},
  author={Forsgren, Seth and Martiros, Hayk},
  journal={URL https://riffusion.com},
  year={2022}
}

@article{mubert2022mubert,
  title={Mubert},
  author={Mubert-Inc.},
  journal={URL https://mubert.com/, https://github.com/MubertAI/Mubert-Text-to-Music},
  year={2022}
}

@article{agostinelli2023musiclm,
  title={MusicLM: Generating music from text},
  author={Agostinelli et al., Andrea},
  journal={arXiv preprint arXiv:2301.11325},
}

@article{schneider2023mo,
  title={Mo$\backslash$\^{}usai: Text-to-music generation with long-context latent diffusion},
  author={Schneider et al., Flavio},
  journal={arXiv preprint arXiv:2301.11757},
}

@article{lam2023efficient,
  title={Efficient neural music generation},
  author={Lam et al., Max WY},
  journal={NeurIPS},
  volume={36},
  pages={17450--17463},
  year={2023}
}

@article{copet2023simple,
  title={Simple and controllable music generation},
  author={Copet et al., Jade},
  journal={NeurIPS},
  volume={36},
  pages={47704--47720},
  year={2023}
}

@inproceedings{li2024jen,
  title={Jen-1: Text-guided universal music generation with omnidirectional diffusion models},
  author={Li et al., Peike Patrick},
  booktitle={CAI},
  pages={762--769},
  year={2024},
  organization={IEEE}
}

@inproceedings{karras2024analyzing,
  title={Analyzing and improving the training dynamics of diffusion models},
  author={Karras et al., Tero},
  booktitle={CVPR},
  pages={24174--24184},
  year={2024}
}

@inproceedings{hong2023improving,
  title={Improving sample quality of diffusion models using self-attention guidance},
  author={Hong et al., Susung},
  booktitle={ICCV},
  pages={7462--7471},
  year={2023}
}

@article{li2025self,
  title={Self-guidance: Boosting flow and diffusion generation on their own},
  author={Li et al., Tiancheng},
  journal={TPAMI},
  year={2025},
  publisher={IEEE}
}

@article{sadat2024no,
  title={No training, no problem: Rethinking classifier-free guidance for diffusion models},
  author={Sadat et al., Seyedmorteza},
  journal={arXiv preprint arXiv:2407.02687},
}

@inproceedings{jiang2025freeaudio,
  title={Freeaudio: Training-free timing planning for controllable long-form text-to-audio generation},
  author={Jiang, Yuxuan and others},
  booktitle={ACMMM},
  pages={9871--9880},
  year={2025}
}

@article{jiang2025controlaudio,
  title={ControlAudio: Tackling Text-Guided, Timing-Indicated and Intelligible Audio Generation via Progressive Diffusion Modeling},
  author={Jiang, Yuxuan and others},
  journal={arXiv preprint arXiv:2510.08878},
  year={2025}
}

@article{dai2026omni2sound,
  title={Omni2Sound: Towards Unified Video-Text-to-Audio Generation},
  author={Dai, Yusheng and others},
  journal={arXiv preprint arXiv:2601.02731},
  year={2026}
}

@article{dai2025latent,
  title={Latent swap joint diffusion for long-form audio generation},
  author={Dai, Yusheng and others},
  journal={arXiv preprint arXiv:2502.05130},
  year={2025}
}
}

\newpage
\appendices

\section{Proof of HG}
\label{proof}
\begin{theorem}
Embedding AG into CFG is equivalent to embedding CFG into AG. Namely, 
\begin{equation}
\label{eq:CFG-AG}
\begin{aligned}
&\bm{\epsilon}_{\text{cAG}}(\bm z_t,t,\bm c)=\bm{\epsilon}_{\theta_{\text{bad}}}(\bm z_t,t,\bm c)+w_1(\bm{\epsilon}_\theta(\bm z_t,t,\bm c)-\bm{\epsilon}_{\theta_{\text{bad}}}(\bm z_t,t,\bm c))\\
&\bm{\epsilon}_{\text{ucAG}}(\bm z_t,t)=\bm{\epsilon}_{\theta_{\text{bad}}}(\bm z_t,t)+w_2(\bm{\epsilon}_\theta(\bm z_t,t)-\bm{\epsilon}_{\theta_{\text{bad}}}(\bm z_t,t)) \\
&\bm{\epsilon}_{\text{HG}}(\bm z_t,t,\bm c)=\bm{\epsilon}_{\text{ucAG}}(\bm z_t,t)+w_3(\bm{\epsilon}_{\text{cAG}}(\bm z_t,t,\bm c)-\bm{\epsilon}_{\text{ucAG}}(\bm z_t,t))
\end{aligned}
\end{equation}
and
\begin{equation}
\label{eq:AG-CFG}
\begin{aligned}
&\bm{\epsilon}_{\text{CFG}}(\bm z_t,t, \bm c)=\bm{\epsilon}_{\theta}(\bm z_t,t)+\hat w_1(\bm{\epsilon}_\theta(\bm z_t,t, \bm c)-\bm{\epsilon}_{\theta}(\bm z_t,t))\\
&\bm{\epsilon}_{\text{badCFG}}(\bm z_t,t, \bm c)=\bm{\epsilon}_{\theta_{\text{bad}}}(\bm z_t,t)+\hat w_2(\bm{\epsilon}_{\theta_{\text{bad}}}(\bm z_t,t, \bm c)-\bm{\epsilon}_{\theta_{\text{bad}}}(\bm z_t,t)) \\
&\hat{\bm{\epsilon}}(\bm z_t,t, \bm c)=\bm{\epsilon}_{\text{badCFG}}(\bm z_t,t, \bm c)+\hat w_3(\bm{\epsilon}_{\text{CFG}}(\bm z_t,t, \bm c)-\bm{\epsilon}_{\text{badCFG}}(\bm z_t,t, \bm c))
\end{aligned}
\end{equation}
are equivalent guidance, as long as $w_3 \notin \{0,1\}$ and $\hat w_3 \notin \{0,1\}$ (HG degenerates to CFG-only or AG-only in these cases) \textit{i.e.}, for any fixed $\bm{\epsilon}_\theta$ and $\bm{\epsilon}_{\theta_\text{bad}}$ ,
\begin{equation}
\label{w_set}
\begin{aligned}
&\{\bm{\epsilon} | \exists w_1,w_2 \in \mathbb{R},w_3 \notin \{0,1\}, \text{s.t.} \bm{\epsilon}=\bm{\epsilon}_\text{HG}(\bm z_t,t, \bm c)\} \\
=&\{\bm{\epsilon} | \exists \hat w_1, \hat w_2 \in \mathbb{R},\hat w_3 \notin \{0,1\}, \text{s.t.} \bm{\epsilon}=\hat {\bm{\epsilon}}_\text{HG}(\bm z_t,t, \bm c)\}.
\end{aligned}
\end{equation}
\end{theorem}
\noindent \textbf{Proof}. Let the two sets in~\eqref{w_set} be denoted as $S_1$ and $S_2$. Substituting the first two equations of~\eqref{eq:CFG-AG} into the third gives:
\begin{equation}
\label{eq:CFG-AG-expanded}
\begin{aligned}
&\bm{\epsilon}_\text{HG}(\bm z_t,t, \bm c)=w_3 \bm{\epsilon}_{\text{cAG}}(\bm z_t,t, \bm c)+(1-w_3)\bm{\epsilon}_\text{ucAG}(\bm z_t,t, \bm c) \\
&=w_3\left(w_1\bm{\epsilon}_\theta(\bm z_t,t, \bm c)+(1-w_1)\bm{\epsilon}_{\theta_\text{bad}}(\bm z_t,t, \bm c)\right) \\
&+(1-w_3)\left(w_2\bm{\epsilon}_\theta(\bm z_t,t)+(1-w_2)\bm{\epsilon}_{\theta_\text{bad}}(\bm z_t,t)\right) \\
&=w_1w_3 \bm{\epsilon}_\theta(\bm z_t,t, \bm c)+w_2(1-w_3) \bm{\epsilon}_\theta(\bm z_t,t) \\
&+w_3(1-w_1) \bm{\epsilon}_{\theta_\text{bad}}(\bm z_t,t, \bm c)+(1-w_2)(1-w_3)\bm{\epsilon}_{\theta_\text{bad}}(\bm z_t,t).
\end{aligned}
\end{equation}
Likewise, Equation~\eqref{eq:AG-CFG} gives
\begin{equation}
\label{eq:AG-CFG-expanded}
\begin{aligned}
&\hat{\bm{\epsilon}}_\text{HG}(\bm z_t,t, \bm c)=\hat w_3 \bm{\epsilon}_{\text{CFG}}(\bm z_t,t, \bm c)+(1-\hat w_3)\bm{\epsilon}_\text{badCFG}(\bm z_t,t, \bm c) \\
&=\hat w_3\left(\hat w_1\bm{\epsilon}_\theta(\bm z_t,t, \bm c)+(1-\hat w_1)\bm{\epsilon}_{\theta}(\bm z_t,t)\right) \\
&+(1-\hat w_3)\left(\hat w_2\bm{\epsilon}_{\theta_\text{bad}}(\bm z_t,t, \bm c)+(1-\hat w_2)\bm{\epsilon}_{\theta_\text{bad}}(\bm z_t,t)\right) \\
&=\hat w_1\hat w_3 \bm{\epsilon}_\theta(\bm z_t,t, \bm c)+\hat w_3(1-\hat w_1) \bm{\epsilon}_\theta(\bm z_t,t) \\
&+\hat w_2(1-\hat w_3) \bm{\epsilon}_{\theta_\text{bad}}(\bm z_t,t, \bm c)+(1-\hat w_2)(1-\hat w_3)\bm{\epsilon}_{\theta_\text{bad}}(\bm z_t,t).
\end{aligned}
\end{equation}
\noindent For most cases $\bm{\epsilon}_\theta(\bm z_t,t)$, $\bm{\epsilon}_\theta(\bm z_t,t, \bm c)$, $\bm{\epsilon}_{\theta_\text{bad}}(\bm z_t,t)$ and $\bm{\epsilon}_{\theta_\text{bad}}(\bm z_t,t, \bm c)$ are linearly independent, thus Equations~\eqref{eq:CFG-AG-expanded} and~\eqref{eq:AG-CFG-expanded} imply
\begin{equation}
\label{eq:coefficients_correspond}
\left\{
\begin{aligned}
w_1w_3 &= \hat w_1 \hat w_3 \\
w_2(1-w_3) &= \hat w_3 (1- \hat w_1) \\
w_3(1-w_1) &= \hat w_2 (1- \hat w_3) \\
(1-w_2)(1-w_3) &= (1-\hat w_2)(1- \hat w_3)
\end{aligned}
\right.
\end{equation}
For Equations~\eqref{eq:coefficients_correspond}, adding the first and third equations yields
\begin{equation}
w_3=\hat w_1 \hat w_3+\hat w_2(1-\hat w_3).
\end{equation}
Substitute into the first and second equations and we can figure out (This division is valid since $w_3 \notin \{0,1\}$)
\begin{equation}
\label{eq:hat-to-w1-w2}
\begin{aligned}
w_1 &= \frac{\hat w_1 \hat w_3}{\hat w_1 \hat w_3+\hat w_2(1-\hat w_3)} \\
w_2 &= \frac{\hat w_3 (1-\hat w_1)}{1-\hat w_1 \hat w_3 - \hat w_2(1- \hat w_3)}
\end{aligned}
\end{equation}
Therefore, $S_2 \subseteq S_1$. Notice that Equations~\eqref{eq:coefficients_correspond} is symmetric under the exchange of $w_1$ and $\hat w_1$, $w_2$ and $\hat w_2$, $w_3$ and $\hat w_3$, so similarly
\begin{equation}
\begin{aligned}
\hat w_1 &= \frac{w_1 w_3}{w_1 w_3+w_2(1-w_3)} \\
\hat w_2 &= \frac{w_3 (1-w_1)}{1-w_1 w_3 -  w_2(1- w_3)} \\
\hat w_3 &=w_1w_3+w_2(1-w_3)
\end{aligned}
\end{equation}
 deducing that $S_1 \subseteq S_2$. Then we have $S_1=S_2$, which completes the proof.

\begin{figure*}[t]
  \centering
  \begin{subfigure}[t]{0.32\textwidth}
    \centering
    \includegraphics[width=1.05\linewidth]{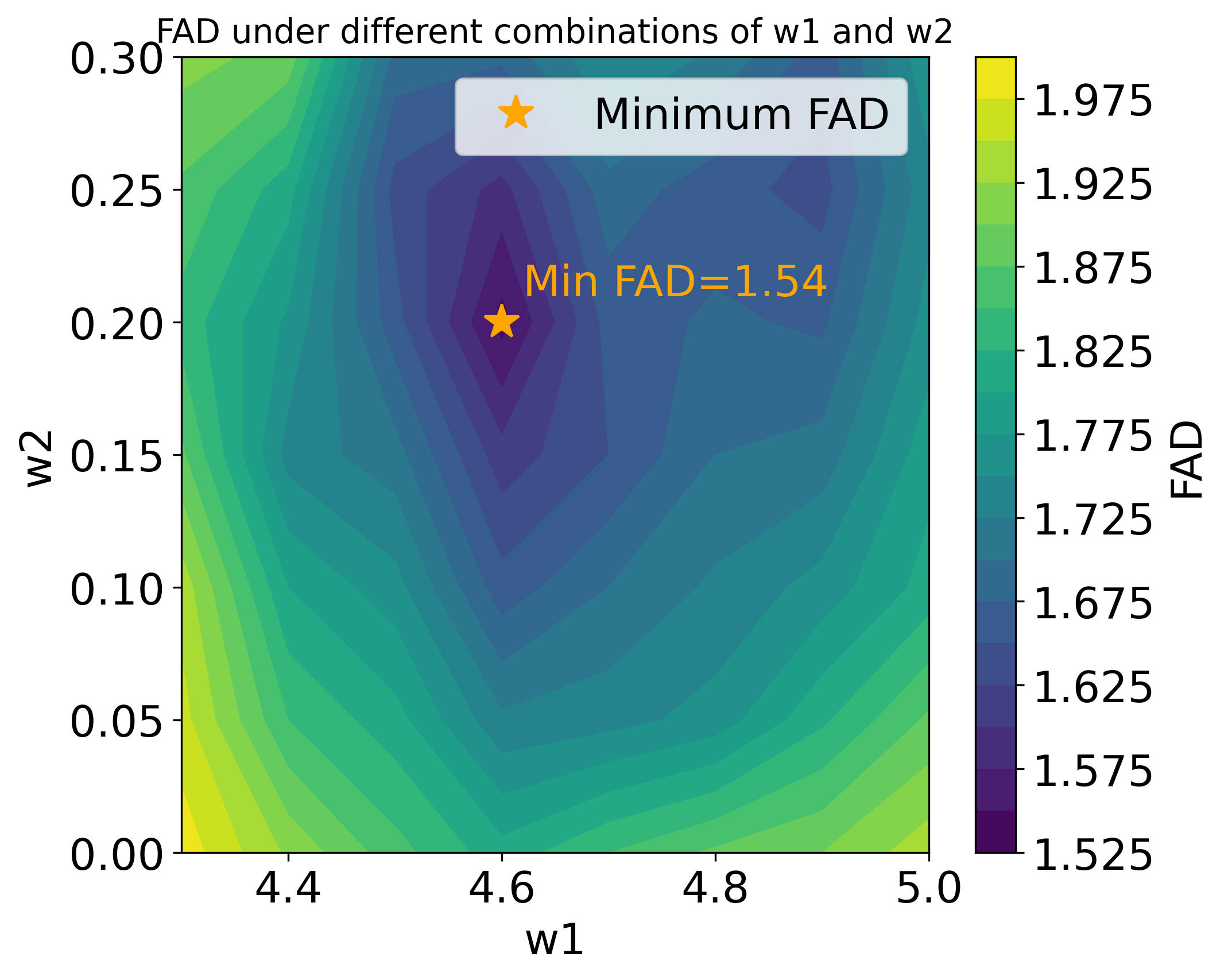}
    \caption{FAD w.r.t. $w_1$ and $w_2$}
    \label{fig:pg_fad}
  \end{subfigure}
  \hfill
  \begin{subfigure}[t]{0.32\textwidth}
    \centering
    \includegraphics[width=1.05\linewidth]{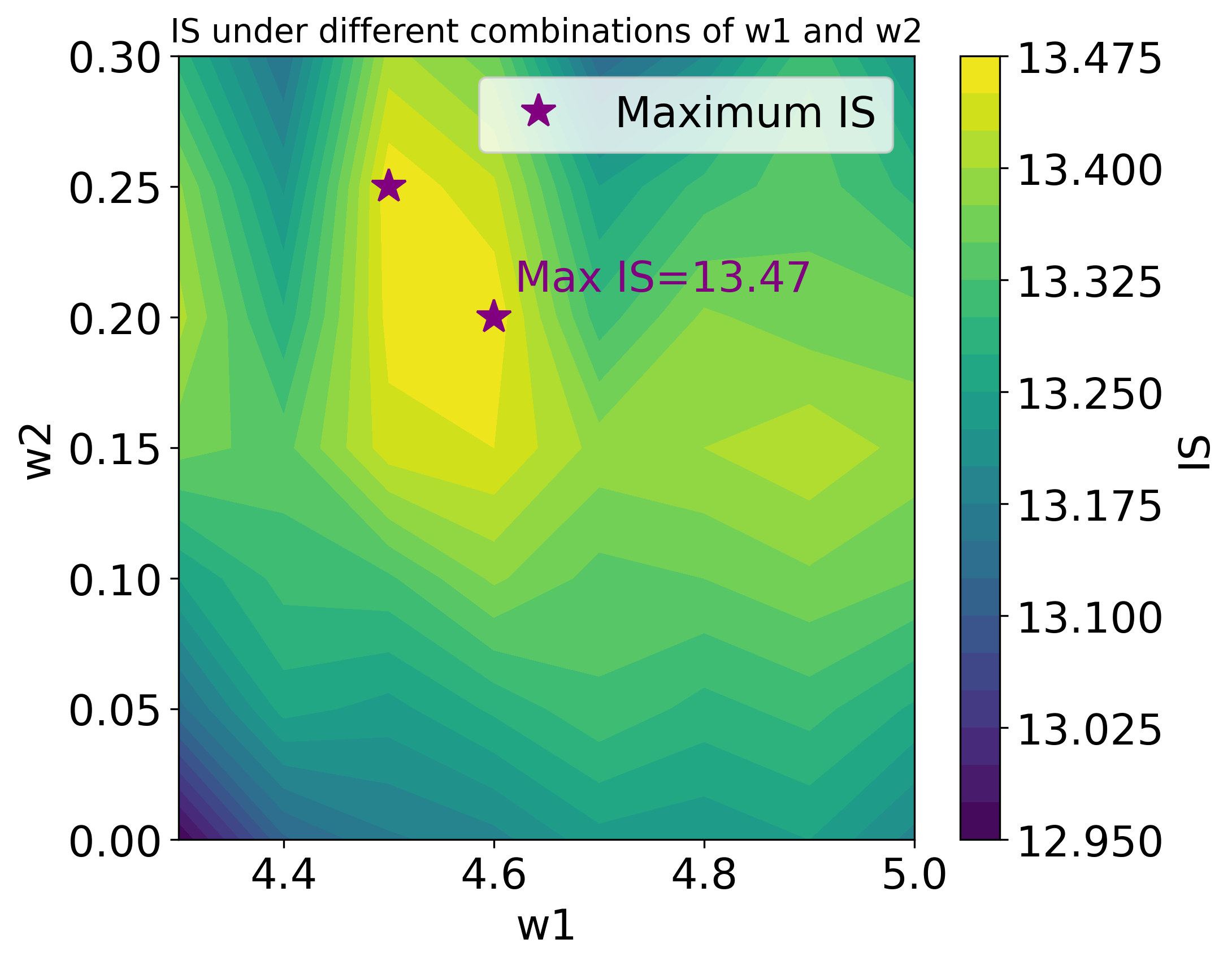}
    \caption{IS w.r.t. $w_1$ and $w_2$}
    \label{fig:pg_is}
  \end{subfigure}
  \hfill
  \begin{subfigure}[t]{0.32\textwidth}
    \centering
    \includegraphics[width=1.05\linewidth]{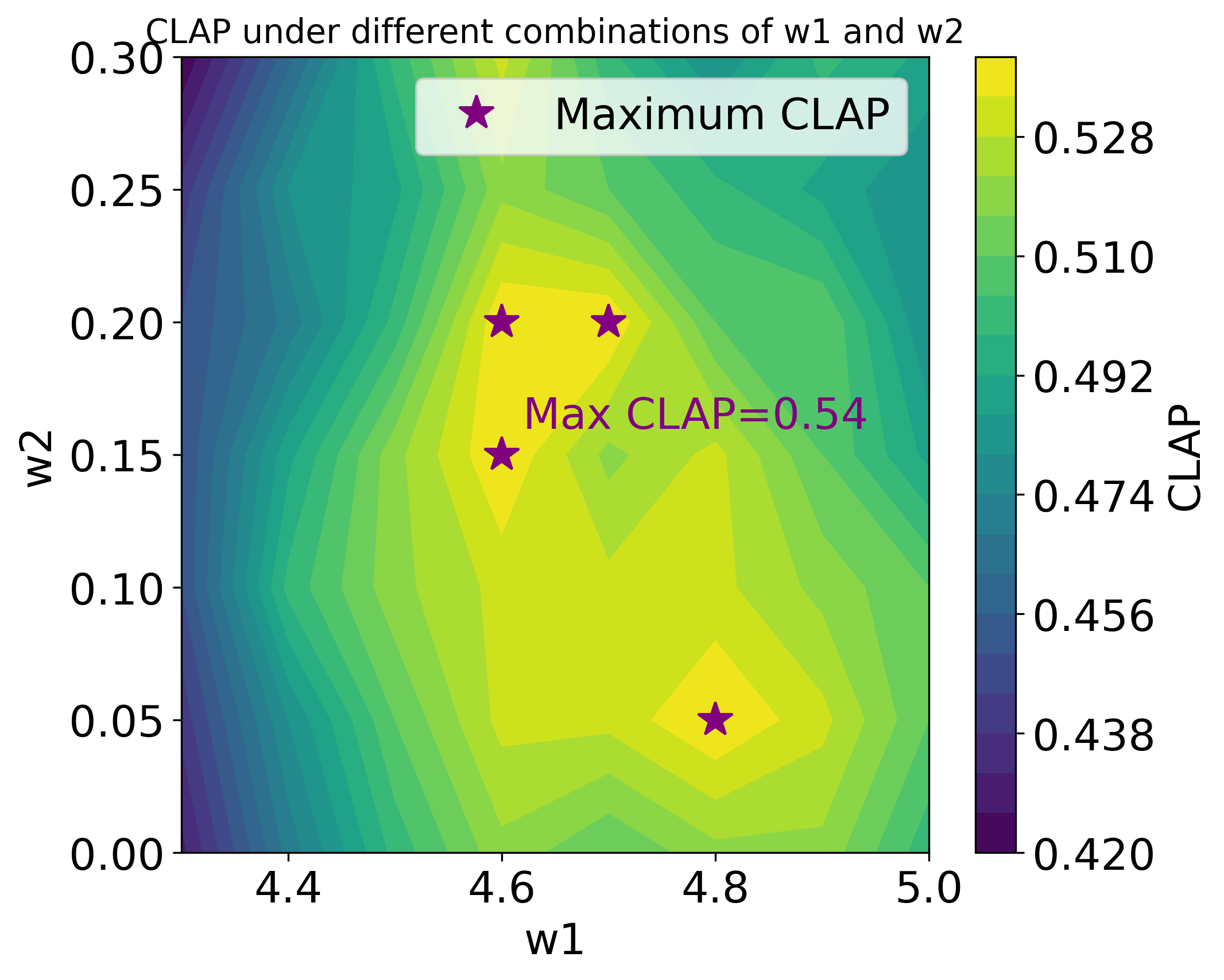}
    \caption{CLAP w.r.t. $w_1$ and $w_2$}
    \label{fig:pg_clap}
  \end{subfigure}
  \caption{Impact of different guidance scales in PG. \textbf{(a)(b)(c)} stands for the relations of FAD, IS and CLAP with $w_1$ and $w_2$, respectively. The best configurations are denoted with stars.}
  \label{fig:pg_weight_comparison}
  \vspace{-0.15in}
\end{figure*}

\begin{figure*}[t]
  \centering
  \begin{subfigure}[t]{0.32\textwidth}
    \centering
    \includegraphics[width=1.05\linewidth]{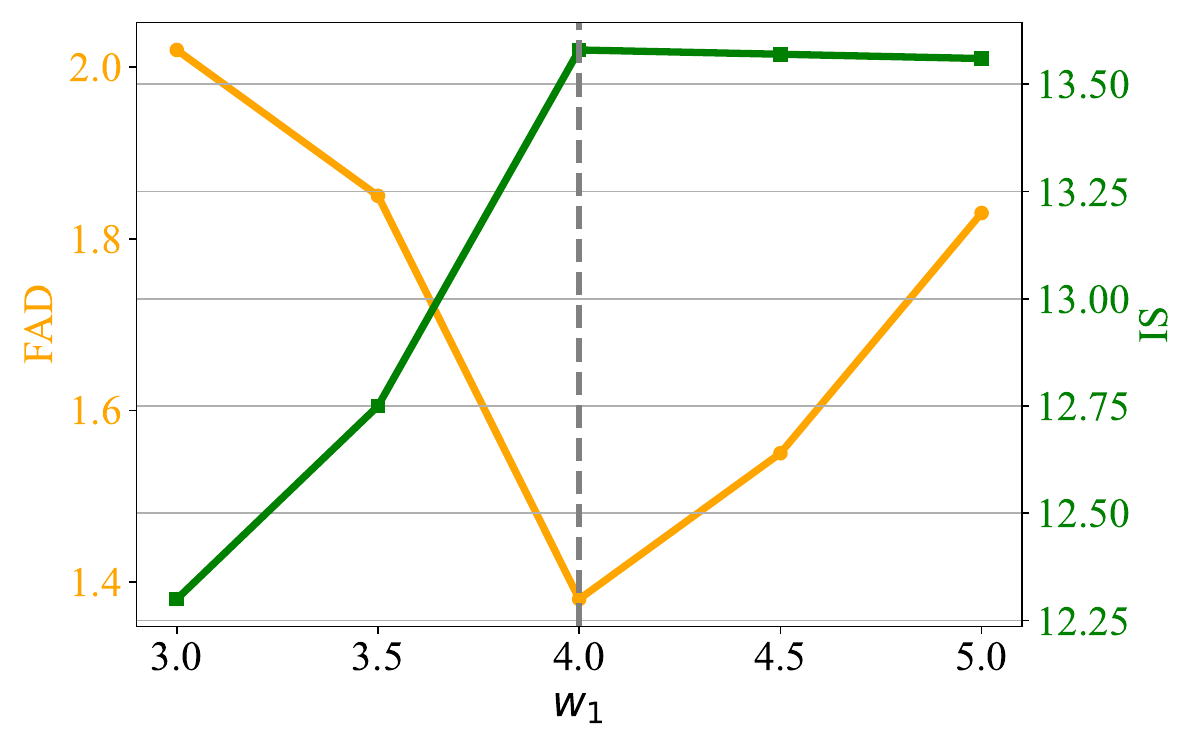}
    \caption{FAD and IS w.r.t. $w_1$}
    \label{fig:w1}
  \end{subfigure}
  \hfill
  \begin{subfigure}[t]{0.32\textwidth}
    \centering
    \includegraphics[width=1.05\linewidth]{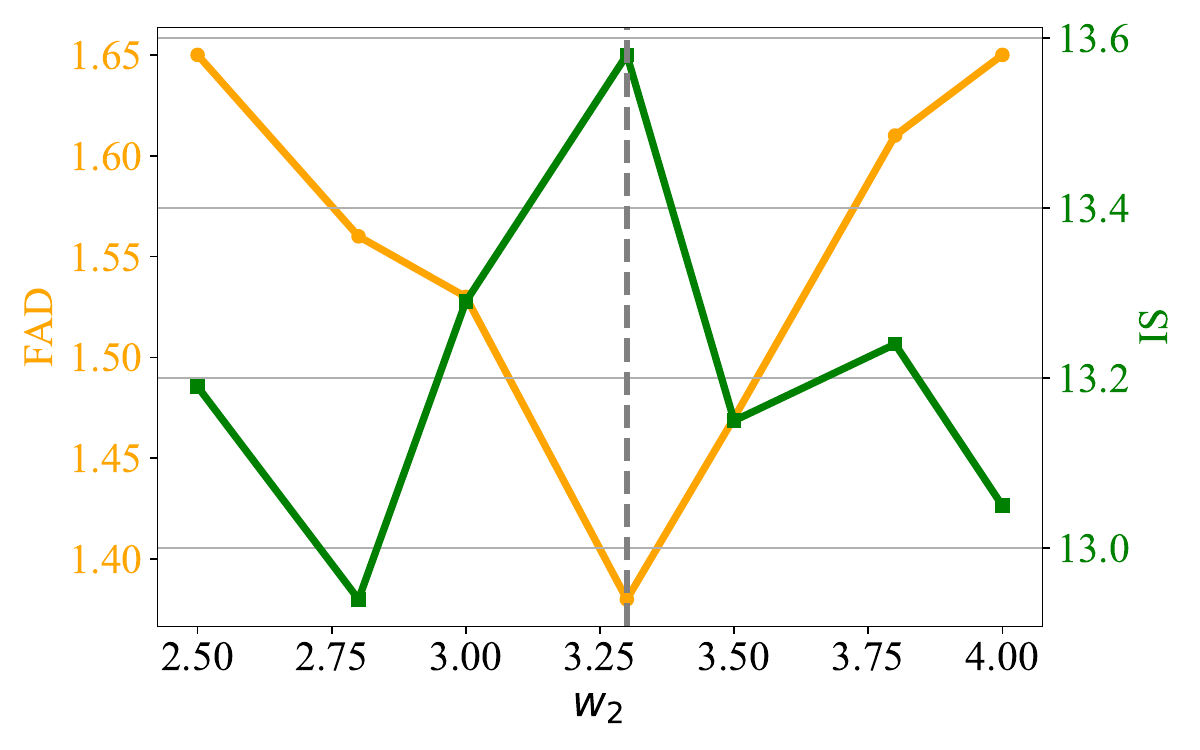}
    \caption{FAD and IS w.r.t. $w_2$}
    \label{fig:w2}
  \end{subfigure}
  \hfill
  \begin{subfigure}[t]{0.32\textwidth}
    \centering
    \includegraphics[width=1.05\linewidth]{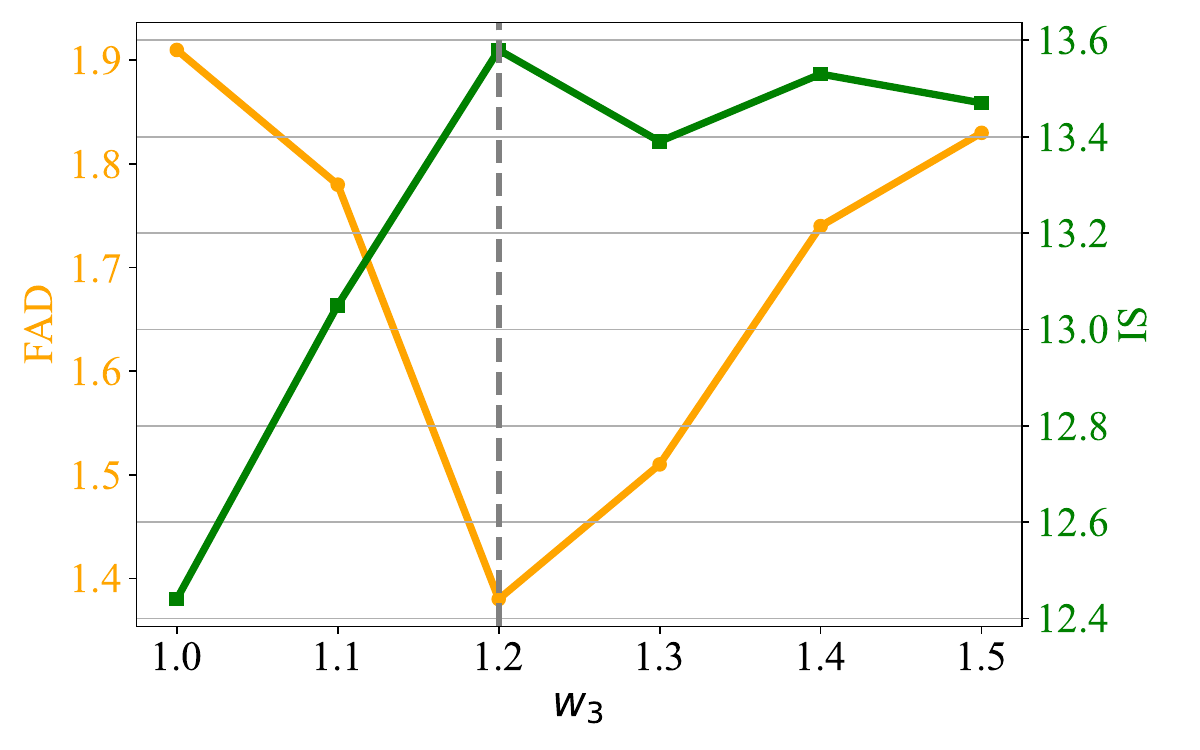}
    \caption{FAD and IS w.r.t. $w_3$}
    \label{fig:w3}
  \end{subfigure}
  \caption{Impact of different guidance scales in HG. \textbf{(a)} Sweep over $w_1$ while keeping $w_2$ and $w_3$ unchanged. \textbf{(b)} Sweep over $w_2$ while keeping $w_1$ and $w_3$ unchanged. \textbf{(c)} Sweep over $w_3$ while keeping $w_1$ and $w_2$ unchanged. The best configurations are denoted with dashed lines.}
  \label{fig:weight_comparison}
  \vspace{-0.15in}
\end{figure*}

\section{Impact of guidance scales}
\label{scales}
To further demonstrate the effectiveness of AudioMoG, we investigate the impact of different guidance scales in PG and HG on the generation results.

\textbf{The PG setting.} We study the guidance scales
$w_1$ in the range $4.3 \sim 5.0$ and $w_2$ in $0.0 \sim 0.3$. We keep the NFE fixed at 300 following the main paper and evaluated FAD, IS and CLAP. The results are shown in Fig. \ref{fig:pg_fad}-\ref{fig:pg_clap}. Noticing that PG degrades to CFG-only when $w_2=0$, we can see that PG indeed reaches a higher performance ceiling and achieves the best results around our choice, $w_1=4.6$ and $w_2=0.2$. Excessively increasing or decreasing $w_1$ and $w_2$ will result in a loss of quality, showing alike convex pattern in the following HG setting.

\textbf{The HG setting.} We adopt the setting in the main paper, with the baseline configuration of $w_1 = 4.0$, $w_2 = 3.3$, and $w_3 = 1.2$, while keeping the NFE fixed at 400. We evaluated both FAD and IS as primary metrics, and the results are summarized in Fig. \ref{fig:w1}-\ref{fig:w3}. For $w_1$ (Fig. \ref{fig:w1}), we can see as $w_1$ increases from 3.0 to 4.0, FAD decreases substantially, reaching a minimum around $w_1 = 4.0$, indicating improved overall sample fidelity. However, further increasing $w_1$ beyond this point slightly worsens FAD. Meanwhile, IS increases steadily and saturates at $w_1 \geq 4.0$, reflecting stronger condition adherence and sample specificity. For $w_2$ and $w_3$, we observe similar convex patterns (Fig. \ref{fig:w2}, \ref{fig:w3}). Both FAD and IS achieve their best values around $w_2=3.3$ and $w_3=1.2$. Therefore, all three scales exhibit non-monotonic behaviors, and each has an optimal value range where both fidelity and diversity metrics are simultaneously optimized. These results validate the necessity of tuning each guidance component, as also noted in \cite{karras2024guiding}, and highlight that balanced guidance from multiple perspectives is crucial for high-quality audio generation.

\section{Text-to-music generation results}
\label{appt2m}

\begin{table}[htbp]
\centering
  \caption{Objective metrics for text-to-music generation on MusicCaps dataset.}
  \label{tab:obj_metrics_music}
  \begin{tabular}{l | c c c}
\toprule
\textbf{Model} & \textbf{FAD $\downarrow$} & \textbf{IS $\uparrow$} & \textbf{FD $\downarrow$}  \\
\midrule
Riffusion \cite{forsgren2022riffusion} & 13.40 & / & /   \\
Mubert \cite{mubert2022mubert} & 9.60 & / & / \\
MusicLM \cite{agostinelli2023musiclm} & 4.00 & / & /  \\
Mousai \cite{schneider2023mo} & 7.50 & / & / \\
MeLoDy \cite{lam2023efficient} & 5.41 & / & /\\
Stable Audio Open \cite{evans2025stable} & 3.51 & 2.93 & 36.42 \\
MusicGen w/o melody \cite{copet2023simple} & 3.40 & / & / \\
AudioLDM 2-Large \cite{liu2024audioldm} & 2.93 & 2.59 & 16.34 \\
AudioLDM 2-Full \cite{liu2024audioldm} & 3.13 & / & /  \\
TANGO-AF \cite{kong2024improving} & 2.21 & 2.79 & 22.69  \\
Jen-1 \cite{li2024jen} & 2.00 & / & / \\
\midrule
CFG-only, $w=7$ & 2.36 & 3.10 & \underline{14.39} \\
MoG-PG, $w_1=2.0,w_2=0.2$ & \underline{1.98} & \underline{4.35} & 14.88 \\
MoG-HG, $w_1=1.6,w_2=1.6,w_3=1.2$ & \textbf{1.92} & \textbf{4.39} & \textbf{14.09}  \\
\bottomrule
\end{tabular}
\end{table}

\subsection{Baseline methods}
To present comprehensive evaluation results, we introduce 10 text-to-music (T2M) baseline models for comparison:

\textbf{Riffusion} \cite{forsgren2022riffusion} is a unique model that generates music by converting spectrogram images into audio. It fine-tunes the Stable Diffusion model on spectrograms, allowing it to produce short music loops based on text prompts.

\textbf{Mubert} \cite{mubert2022mubert} is an AI-driven music generation platform that creates royalty-free music tailored for various content needs. It offers tools for content creators, artists, and developers to generate and integrate AI-generated music into their projects. 

\textbf{MusicLM} \cite{agostinelli2023musiclm} is a model introduced by Google that generates high-fidelity music from text descriptions. It utilizes a sequence-to-sequence modeling approach to capture long-term structure in music generation. 

\textbf{Mousai} \cite{schneider2023mo} is a two-stage latent-diffusion system for text-to-music generation. Stage 1 compresses 48 kHz stereo audio with a Diffusion-Magnitude Autoencoder (DMAE), and Stage 2 is a text-conditioned latent diffusion (TCLD) model that can produce multi-minute, prompt-aligned musical pieces.

\textbf{MeLoDy} \cite{lam2023efficient} is an LM-guided diffusion framework that inherits
 the highest-level LM from MusicLM for semantic modeling, and applies a novel
 dual-path diffusion (DPD) model and an audio VAE-GAN to efficiently decode the
 conditioning semantic tokens into waveform, resulting in cutting sampling cost by more than 95 \% while maintaining state-of-the-art text–music alignment and audio quality.

\textbf{MusicGen} \cite{copet2023simple} is an open-source model developed by Meta that generates music from text prompts. It employs a transformer-based architecture trained on a large dataset of music to produce diverse and high-quality audio samples. 

\textbf{Jen-1} \cite{li2024jen} is a universal high-fidelity model for text-to-music generation. It incorporates both autoregressive and non-autoregressive training, enabling tasks like text-guided music generation, inpainting, and continuation.

\subsection{Experiment results}
We conduct a comprehensive analysis of generated music quality across the above systems. For Riffusion, Mubert and MusicLM, we report the metrics from MusicLM. For Stable Audio Open, AudioLDM2-Large and Tango-AF, we cite the results from ETTA. For Mousai, as it is not evaluated in ETTA, we report it from AudioLDM 2. For other baselines, we report the metrics as presented in their original papers. Similar to T2A, we further evaluate our baseline model using only CFG, which also achieves the best result when $w=7$, and fix NFE to 400. All evaluations are conducted on MusicCaps dataset. The results are detailed in Table \ref{tab:obj_metrics_music}. It can be shown that HG not only surpasses CFG-only but also achieves SOTA in all metrics.




\section{Image generation results}
\label{appt2i}
This section presents the experimental results for the conditional ImageNet $512 \times 512$ generation task, using the EDM2-S checkpoints from \cite{karras2024analyzing}. Image generation was performed over 16 deterministic steps with a second-order Heun sampler, maintaining the same inference speed with a single guidance method~\cite{karras2024guiding}. The optimal guidance strengths were determined through a grid search on a reduced sample size ($N=8192$). Subsequently, the model's performance was evaluated on a larger set of samples ($N=50000$) to obtain robust estimates of the Fréchet Inception Distance (FID) and the Fréchet DINOv2 Distance (FD$_{\textnormal{DINOv2}}$).

\begin{table}[htbp]
\centering
  \caption{Conditional image generation results on ImageNet-512.}
  \label{tab:imagenet}
  \begin{tabular}{l | c c c}
\toprule
\textbf{Model} & \textbf{FID $\downarrow$} & \textbf{FD$_{\textnormal{DINOv2}}$ $\downarrow$} \\
\midrule
CFG-only & 2.40 & 96.90  \\
AG-only & 1.60 & 57.35  \\
MoG-PG & 1.60 & 53.01  \\
MoG-HG & \textbf{1.47} & \textbf{49.92}   \\
\bottomrule
\end{tabular}
\end{table}

\section{2D toy dataset experiment details}
\label{2DtoyDetail}

This section outlines the setup of the 2D toy dataset experiment used in the analysis presented in Section~\ref{AnalysisofGuidance}. Unless otherwise specified, the experiment details strictly follow the setup in~\cite{karras2024guiding}.

\textbf{Dataset.}
The dataset is a synthetic, fractal-like 2D distribution composed of two classes. Each class is represented as a Gaussian mixture model $\mathcal{M}_c = (\phi_i, \mu_i, \bm{\Sigma}_i)$, where $\phi_i$ denotes the mixture weight, $\mu_i$ the mean, and $\bm{\Sigma}_i$ the $2\times2$ covariance matrix of the $i$-th Gaussian component.

\textbf{Models.}
We employ simple multi-layer perceptrons (MLPs) as denoiser models, consistent with the setup in~\cite{karras2024guiding}.

\textbf{Training.}
To ensure comparability, we use the pre-trained models provided by~\cite{karras2024guiding}. The URLs of model checkpoints can be found at: \url{https://github.com/NVlabs/edm2/blob/main/toy_example.py}.

\textbf{Sampling.}
For the visualizations in Fig.~\ref{fig:toy}, we use the following guidance weights: $w=3$ for both CFG and AG; $w_1=2, w_2=2$ for PG; and $w_1=w_2=1.5$, $w_3=2$ for HG. The models $\bm{\epsilon}_\theta(\bm z_t, t, \bm c)$ and $\bm{\epsilon}_\theta(\bm z_t, t)$ use checkpoints trained with hidden dimension $d=64$ and training iteration $M=4096$, while the bad models $\bm{\epsilon}_{\theta_{\text{bad}}}(\bm z_t, t, \bm c)$ and $\bm{\epsilon}_{\theta_{\text{bad}}}(\bm z_t, t)$ use checkpoints with $d=32$ and $M=512$.

\section{Text-to-audio experiment details}
\label{experimentdetails}

\subsection{Datasets}
\label{datasets}
We present the datasets used to train our baseline model in Table \ref{tab:obj_metrics}. AudioCaps \cite{kim2019audiocaps} is a benchmark dataset for audio captioning that contains 50,000 audio clips from AudioSet paired with human-written textual descriptions. We only used its training set. AudioSet \cite{gemmeke2017audio} is a large-scale weakly labeled dataset released by Google, comprising over 2 million 10-second audio clips across more than 600 sound event categories. The BBC Sound Effects Library 
[\href{sound-effects.bbcrewind.co.uk}{link}] provides over 30,000 professionally recorded sound effects covering a wide variety of acoustic scenes and events. Clotho v2 \cite{drossos2020clotho} is designed for audio captioning tasks and contains approximately 5,000 audio clips, each annotated with five crowdsourced textual descriptions. VGGSound \cite{chen2020vggsound} is a large-scale dataset consisting of over 200,000 10-second video clips from YouTube, covering 310 diverse sound classes. FreeSound [\href{freesound.org}{link}] is a collaborative platform that hosts a wide range of user-contributed audio samples, frequently used for environmental sound classification and retrieval. FSD50K \cite{fonseca2021fsd50k} is a large-scale dataset derived from FreeSound, containing over 50,000 audio clips annotated with strong and weak labels for sound event detection. FMA \cite{defferrard2016fma}, the Free Music Archive dataset, includes full-length high-quality music tracks and is widely used in music information retrieval research. The Million Song Dataset (MSD) \cite{bertin2011million} offers metadata and pre-computed audio features for one million popular music tracks to support large-scale music recommendation and analysis. MagnaTagATune (MTT) \cite{law2009evaluation} is a music tagging dataset that contains 25,000 audio clips annotated with multiple descriptive tags for genre, instrument, and mood classification tasks.

\begin{table}[htbp]
\centering
  \caption{Statistics for the datasets used in the paper.}
  \label{tab:obj_metrics}
  \begin{tabular}{l c l}
\toprule
\textbf{Dataset} & \textbf{Hours} (h)  & \textbf{Source}  \\
\midrule
AudioCaps & 109 & \cite{kim2019audiocaps} \\
AudioSet & 5800 & \cite{gemmeke2017audio}\\
BBC Sound Effects Library & 300 & \href{sound-effects.bbcrewind.co.uk}{link}\\ 
Clotho v2 & 152 & \cite{drossos2020clotho}\\
VGGSound & 550 & \cite{chen2020vggsound} \\
FreeSound & 6246 & \href{freesound.org}{link}\\
FSD50k & 108 & \cite{fonseca2021fsd50k}\\
FMA &  900 & \cite{defferrard2016fma}\\
MSD & 7333 & \cite{bertin2011million}\\
MTT & 200 & \cite{law2009evaluation} \\
\bottomrule
\end{tabular}
\end{table}

\subsection{Baseline methods}
\label{baseline}
We employ 6 strong T2A baseline methods for comparison. 

\textbf{AudioLDM} is a latent diffusion model for text-to-audio (T2A) generation presented by \cite{liu2023audioldm}. It performs the diffusion process in the latent space of a pretrained audio VAE, while conditioning on text embeddings produced by the CLAP text branch. This design enables a T2A training process without paired data and produces audio that is semantically consistent with the input description.

\textbf{AudioLDM2} is an improved version of the AudioLDM model, introduced by \cite{liu2024audioldm}. It incorporates several improvements, including leveraging the Language of Audio (LOA) encoder and finetuning a GPT-2 model to translate any modality to LOA. These improvements result in higher-quality audio generation that better aligns with the input text.

\textbf{AudioGen} is a text-to-audio generation model introduced by \cite{kreuk2022audiogen}. It employs an autoregressive transformer architecture to generate audio samples conditioned on textual descriptions. The model is trained on a large-scale dataset of audio-text pairs, enabling it to produce high-quality audio outputs that align with the given textual input.

\textbf{Make-An-Audio} is a diffusion-based text-to-audio generation model proposed by \cite{huang2023make}. It introduces a pseudo prompt enhancement with the
distill-then-reprogram approach, including a large number of concept compositions by opening up the usage of language-free audios to alleviate data scarcity. Therefore, it enables high-fidelity, prompt-aligned outputs.

\textbf{TANGO-AF\&AC-FT-AC} \cite{kong2024improving} pre-trains the TANGO architecture on the synthetic-caption AF-AudioSet plus AudioCaps, followed by fine-tuning on AudioCaps alone. Leveraging high-quality synthetic captions significantly improves text-to-audio alignment and overall audio realism.

\textbf{Stable Audio Open} \cite{evans2025stable} is an open-source text-to-audio generation model developed by Stability AI. It leverages a diffusion-based architecture trained on a diverse dataset of audio-text pairs. The model is designed to generate high-fidelity audio samples conditioned on textual input, supporting various applications such as music generation, sound effect synthesis, and more.

\subsection{Model configurations}
\label{config}
Our diffusion model is built upon the Diffusion Transformer (DiT)  architecture, following a latent diffusion modeling (LDM) paradigm that offers strong generative capabilities and effective context modeling. We use FLAN-T5 as the text encoder for our model, and we train a Variational Autoencoder (VAE) that compresses the original waveform into the latent representation. The backbone of the diffusion network adopts a DiT structure with 24 layers and 24 attention heads, each with an embedding dimension of 1536. The model supports both cross-attention and global conditioning: cross-attention is applied to all types of conditional inputs, while global conditioning specifically handles duration-related control signals. The internal token dimension of the diffusion model is set to 64, with a conditional token dimension of 768 and a global condition embedding dimension of 1536. The generated latent representation has the same dimensionality as \texttt{io\_channels}, which is 64.

\subsection{Compression networks}

To train the audio autoencoder, we adopt a variational autoencoder (VAE) architecture based on the Oobleck framework \cite{evans2025stable} at a sampling rate of 16kHz. The model is trained from scratch on large-scale publicly available text-audio paired datasets. The encoder and decoder are symmetric, each using a base channel size of 128, with channel multipliers $1, 2, 4, 8, 16$ and strides $2, 2, 4, 4, 10$. The encoder maps the input waveform into a 128-dimensional latent representation, while the decoder reconstructs the waveform from a 64-dimensional latent code. Snake activation is applied throughout the network, and no final tanh activation is used in the decoder. The overall downsampling ratio is 640, and both input and output are mono-channel waveforms. The bottleneck is implemented as a variational layer.

\subsection{Objective metrics}
\label{objective}
We introduce the objective metrics employed in our evaluation, including Fr\'echet Audio Distance (FAD), Kullback-Leibler (KL) divergence, Inception Score (IS), Fr\'echet Distance (FD), and LAION-CLAP score \cite{wu2023clap}. FAD, adapted from FID (Frechet Inception Distance), measures the distributional gap between generated and reference audio using VGGish embeddings, and serves as our primary indicator of audio fidelity. KL divergence evaluates the difference in acoustic event posteriors between ground truth and generated audio. IS reflects both diversity and specificity of the generated samples, based on entropy over class predictions. FD, while similar in formulation to FAD, is computed in more general embedding spaces and tends to be less stable in audio tasks. We include it for completeness but primarily rely on FAD for fidelity assessment. The CLAP score is calculated as the cosine similarity between CLAP embeddings of the generated audio and the corresponding text. We use the AudioLDM evaluation toolkit to compute all objective metrics.

\subsection{Subjective evaluation}
\label{subjective}
\begin{figure*}[htbp] 
  \centering
  \includegraphics[width=\linewidth]{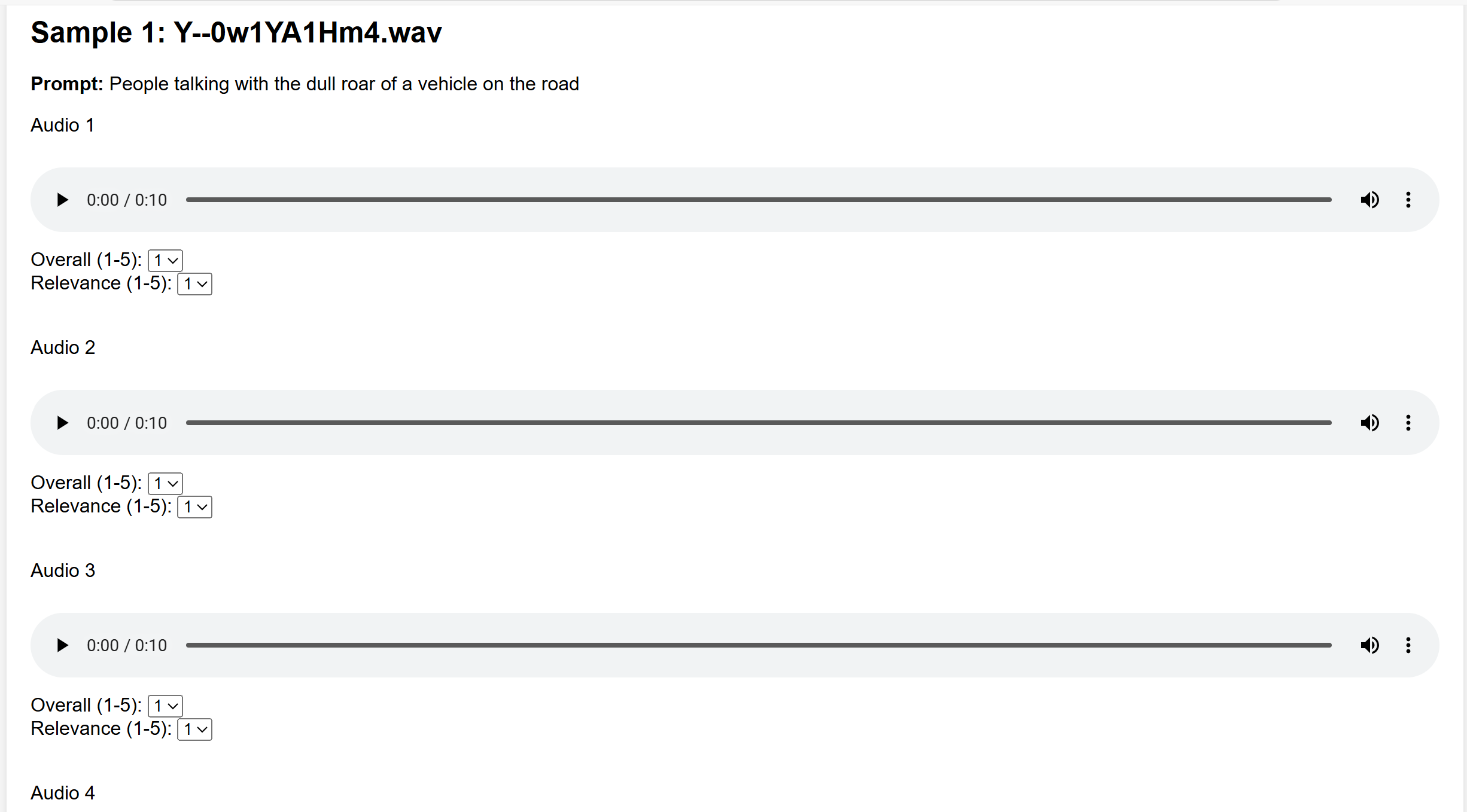}
  \caption{Screenshot of our subjective evaluations.}
  \label{fig:screenshot}
\end{figure*}

We randomly selected 20 samples from the AudioCaps test set for the subjective evaluation. Each group includes the results from AudioLDM, AudioLDM2, CFG-only, HG, and the ground truth (GT), with the order of samples within each group randomly shuffled. Each group was rated by 20 human raters. In our evaluation, both overall quality (OVL) and text relevance (REL) are rated on a scale from 1 to 5. For OVL, raters assess the perceptual quality of the audio, while for REL, they rate the relevance of the audio to the given text condition. The minimum rating increment for all scores is 1 point. A screenshot of our evaluation interface is shown in Fig. \ref{fig:screenshot}.

\section{Video-to-audio experiment details}
\label{appv2a}

\subsection{Datasets}
\label{appv2adatasets}
Apart from T2A and T2M generation, we conduct experiments for video-to-audio (V2A) generation. We utilize the benchmark datasets AudioSet \cite{gemmeke2017audio} and VGGSound \cite{chen2020vggsound} for model training. To compare with previous work, we evaluated our models on the VGGSound test set, which consists of about 15K 10-second audio clips.

\subsection{Model configurations}
\label{appv2asetup}
We use CLIP embeddings to extract the visual features, and we leverage the same VAE as in T2A. The diffusion backbone shares the same core architecture as the T2A counterpart, being built upon the DiT within the LDM paradigm. Most hyperparameters remain the same with T2A, but we increased the conditional token dimension to 1024 to better accommodate high-dimensional visual embeddings, and the global condition embedding dimension is set to 2048. 

We fine-tuned from a T2A model that was trained for 2M iterations with a batch size of 8 per GPU. For the good model, we conducted 1.3M finetuning iterations, while for the bad model, we fine-tuned for 0.3M iterations, both using a batch size of 8 per GPU. The optimizer and sampler settings are the same as those used in the T2A model. In the main results Table~\ref{tab:v2aobjective}, we set $w_1=2.7, w_2=0.15$ in MoG-PG and $w_1=2,7, w_2=2.5, w_3=1.2$ in MoG-HG.

\subsection{Baselines methods}
\label{appv2abaselines}
We introduce 8 V2A baseline models for comparison.

\textbf{IM2WAV} is an image-guided open-domain audio generation system introduced by \cite{sheffer2023hear}. It employs two Transformer language models operating over a discrete audio representation derived from a VQ-VAE model. The system first generates a low-level audio representation using a language model, then upsamples the audio tokens with an additional language model to produce high-fidelity audio. Visual conditioning is achieved through CLIP embeddings, and CFG is applied to steer the generation process.

\textbf{Diff-Foley} is a synchronized V2A synthesis method utilizing a latent diffusion model (LDM), presented by \cite{luo2023diff}. It incorporates contrastive audio-visual pretraining (CAVP) to learn temporally and semantically aligned features, which are then used to train the LDM on spectrogram latent space. The model employs cross-attention modules and "double guidance" to enhance sample quality, achieving state-of-the-art performance in V2A tasks.

\textbf{FoleyGen} is an open-domain V2A generation system based on a language modeling paradigm, introduced by \cite{mei2024foleygen}. It leverages a neural audio codec for bidirectional conversion between waveforms and discrete tokens. A single Transformer model, conditioned on visual features extracted from a visual encoder, facilitates the generation of audio tokens. The model addresses temporal synchronization challenges by exploring novel visual attention mechanisms.

\textbf{VTA-LDM} is a V2A generation framework developed by \cite{xu2024vtaldm}, building upon the LDM framework. It employs a CLIP-based vision encoder to extract frame-level video features, which are temporally concatenated and mapped using a projector as the generation condition. The model focuses on generating semantically and temporally aligned audio content corresponding to video inputs.

\textbf{FoleyCrafter} \cite{zhang2024foleycrafter} is a text-based V2A generation framework designed to produce high-quality, semantically relevant, and temporally synchronized audio for videos. It extends state-of-the-art T2A generators by incorporating a semantic adapter for semantic alignment and a temporal adapter for precise audio-video synchronization, ensuring realistic sound effects that align with visual content.

\textbf{V2A-Mapper} is a lightweight solution for V2A generation proposed by \cite{wang2024v2amapper}. It connects foundation models by translating visual CLIP embeddings into auditory CLAP embeddings, bridging the domain gap between visual and audio modalities. Conditioned on the translated CLAP embedding, a pretrained audio generative model (AudioLDM) is used to produce high-fidelity and visually-aligned sound, requiring minimal training parameters.

\textbf{VAB-Encodec} \cite{su2024vab} is a unified audio-visual framework that learns latent representations and enables vision-to-audio generation within the same model. It tokenizes 48 kHz audio with a pretrained Encodec tokenizer and encodes video frames with an image encoder. During pre-training the model performs visual-conditioned masked-audio-token prediction; at inference it iteratively decodes audio tokens conditioned on visual features, yielding fast and semantically aligned sound.

\textbf{VATT} is a multi-modal generative framework for V2A generation through text, presented by \cite{liu2024VATT}. It comprises two modules: VATT Converter, a large language model fine-tuned for instructions that maps video features to the LLM vector space; and VATT Audio, a transformer that generates audio tokens from visual frames and optional text prompts using iterative parallel decoding. The framework allows for controllable audio generation and audio captioning based on video inputs.

For Diff-foley, VTA-LDM and FoleyCrafter, we generate 10-second audio samples using their
official implementations. For V2A-Mapper, it supplies pre-generated audio samples for evaluation. As the official implementations of FoleyGen, VAB-Encodec and VATT are unavailable, we compare our results with their official  reported results. The IM2WAV results are adopted from VATT.
\subsection{Metrics}
\label{appv2ametrics}
We introduce additional metrics utilized in our V2A evaluation apart from FAD, KL, IS and FD, including Imagebind Score (IBS) and temporal alignment accuracy (AA), which primarily measures audio-visual alignment. 

\textbf{ImageBind Score (IBS)} assesses the semantic alignment between generated audio and the corresponding video by computing the cosine similarity between their embeddings in a shared multimodal space. This metric leverages the ImageBind model \cite{girdhar2023imagebind}, which aligns multiple modalities—including images, audio, and text—into a unified embedding space, facilitating cross-modal retrieval and evaluation . A higher IBS indicates a stronger semantic correlation between the audio and video content.

\textbf{Temporal Alignment Accuracy (AA)} measures the synchronization between generated audio and video by evaluating the model's ability to produce audio events that are temporally aligned with visual events. Introduced in Diff-Foley \cite{luo2023diff}, this metric involves training a classifier to distinguish between correctly aligned audio-video pairs and misaligned ones. The classifier is trained on three types of pairs: true pairs (correctly aligned), temporally shifted pairs, and mismatched pairs from different videos. Align Acc is computed as the percentage of correctly identified true pairs, providing a quantitative measure of temporal synchronization.

By incorporating both IBS and Align Acc, we offer a comprehensive evaluation of the semantic and temporal alignment between generated audio and video, ensuring that the audio not only matches the content but also aligns accurately in time.

\section{Detailed related works}
\label{apprelatedworks}

\subsection{Text-to-audio (T2A) generation}
T2A systems generate audio samples conditioned on natural language prompts. At the beginning, AudioGen~\cite{kreuk2022audiogen} explore autoregressive-based generation methods in the compressed space of mwaveform respectively. Then, DiffSound~\cite{yang2023diffsound}, AudioLDM~\cite{liu2023audioldm} and Make-An-Audio~\cite{huang2023make} develop latent diffusion models in the compressed space of mel-spectrogram, improving overall T2A generation quality. Tango~\cite{deepanway2023text} improves the text encoder of diffusion-based T2A systems with a language model. AudioLDM 2~\cite{liu2024audioldm} employs an autoregressive-based method to predict the AudioMAE features from various input modalities, and then uses a latent diffusion model to generate audio from AudioMAE features. Tango2~\cite{majumder2024tango} and Tangoflux~\cite{hung2024tangoflux} utilize reinforcement strategies including CLAP-guided reward shaping~\cite{wu2023clap} to improve the human preference and semantic-textual alignment.
Recently, Stable Audio~\cite{evans2024fast} designs transformer-based scalable latent diffusion models in the space directly compressed from the audio waveform. ETTA~\cite{lee2024etta} elucidates the design space of diffusion-based T2A systems.

These innovative methods have improved T2A generation quality from generative methods, compression networks, and network architectures,  while the innovations on guidance methods have not been carefully investigated in previous works.

\subsection{Video-to-audio (V2A) generation}
Recent advances in video-to-audio (V2A) generation can be broadly divided into two categories: (1) enhancing V2A via pre-trained text-to-audio (T2A) models, and (2) introducing auxiliary temporal representations to improve temporal alignment. In the first category, methods such as V2A-Mapper \cite{wang2024v2amapper} and FoleyCrafter \cite{mei2024foleygen} build upon established T2A systems like AudioLDM \cite{liu2023audioldm}. These approaches either align video features with the original conditioning space of T2A models or introduce additional adapters to inject visual information as supplementary conditions. For example, V2A-Mapper proposes a mapping strategy that translates video features into audio CLAP embeddings, enabling AudioLDM to perform V2A synthesis. Similarly, FoleyCrafter integrates dedicated adapters to incorporate visual cues into the conditioning process of T2A models.

The second category includes methods such as TiVA \cite{wang2024tiva}, ReWaS \cite{jeong2024rewas}, SyncFusion \cite{comunita2024syncfusion}, and SonicVisionLM \cite{xie2024sonicvisionlm}, which incorporate explicitly designed temporal features to enhance synchronization between video and audio. TiVA employs downsampled Mel spectrograms as auxiliary representations that carry temporal structure, and utilizes a transformer-based predictor to estimate these features for guiding V2A generation. SyncFusion and SonicVisionLM leverage onset positions and audio timestamps, respectively, as temporal control signals during synthesis. ReWaS introduces energy as a continuous temporal representation, providing a more fine-grained condition along the time axis to better regulate V2A output. 
At the sampling stage, CFG is popularly employed in these methods. 
In our paper, we explore a novel sampling algorithm to increase 
generation quality in a training-free and computationally lightweight manner, which is orthogonal to previous innovations.

\subsection{Guidance methods}

\textbf{CFG.}
In previous diffusion-based T2A generation~\cite{liu2023audioldm,liu2024audioldm,huang2023make,huang2023make2,evans2024fast,evans2025stable,lee2024etta,jiang2025freeaudio,jiang2025controlaudio,dai2025latent} and V2A generation works \cite{sheffer2023hear,luo2023diff,mei2024foleygen,xu2024vtaldm,wang2024v2amapper,su2024vab,liu2024VATT,dai2026omni2sound} , CFG~\cite{ho2022classifier} is commonly adopted to improve the audio generation quality at the inference stage. To achieve optimal results, its guidance scale is investigated in different methods~\cite{hung2024tangoflux}. 
However, as demonstrated in recent work~\cite{lee2024etta}, CFG sacrifices the diversity of generation results and may suffer from suboptimal synthesis quality. Previous theoretical analysis~\cite{chidambaram2024does} demonstrates that for any non-zero level of score estimation error, a large CFG strength causes the sampler to diverge from the data distribution's support, formally explaining the empirical phenomenon of distortion at high guidance scales. Our analysis reveals that when CFG uses a bad unconditional model, it inherently introduces a score correction term. This may explain the empirical finding that a small unconditional model can effectively guide a large conditional model~\cite{karras2024analyzing}. However, in standard CFG, this beneficial correction term is entangled with the conditional alignment term, making it difficult to find a guidance strength that simultaneously ensures outlier removal and quality enhancement.

\textbf{Guidance with weak models.}
Recently, AG~\cite{karras2024guiding} proposes a method to guide a diffusion model with the bad version of itself, demonstrating stronger synthesis quality than CFG in image domain. 
Several works~\cite{phunyaphibarn2025unconditional,jeon2025spg,STG} has extended this idea to the scenarios of fine-tuning diffusion models to a specific task~\cite{phunyaphibarn2025unconditional}, motion synthesis~\cite{jeon2025spg}, and video generation~\cite{STG}. Other works adopt similar ideas to construct a weak model. For instance, SAG~\cite{hong2023improving} applies Gaussian blurring to the model input, SG~\cite{li2025self} alters the denoising timestep to obtain an output with higher noise levels, while ICG~\cite{sadat2024no} randomly samples a condition to replace the null embedding in CFG. Among these guidance methods, AG is the most representative. It provides a theoretical justification for its efficacy, positing that it reduces the score estimation error induced by the "mass-covering" or "mean-seeking" behavior of the score matching training objective. However, the advantages of AG have not been observed for audio generation. 
As recently mentioned in ETTA~\cite{lee2024etta}, AG is sensitive to the choice of the bad model, prohibiting their application on audio synthesis.
In this work, we explore the advantages of AG for audio generation, and propose a novel sampling algorithm, MoG, yielding cumulative advantages of AG and CFG to outperform either of them on audio synthesis.

\section{More generated samples}
\label{appgeneratedsamples}
\begin{figure*}[htbp] 
  \centering
  \includegraphics[width=\textwidth]{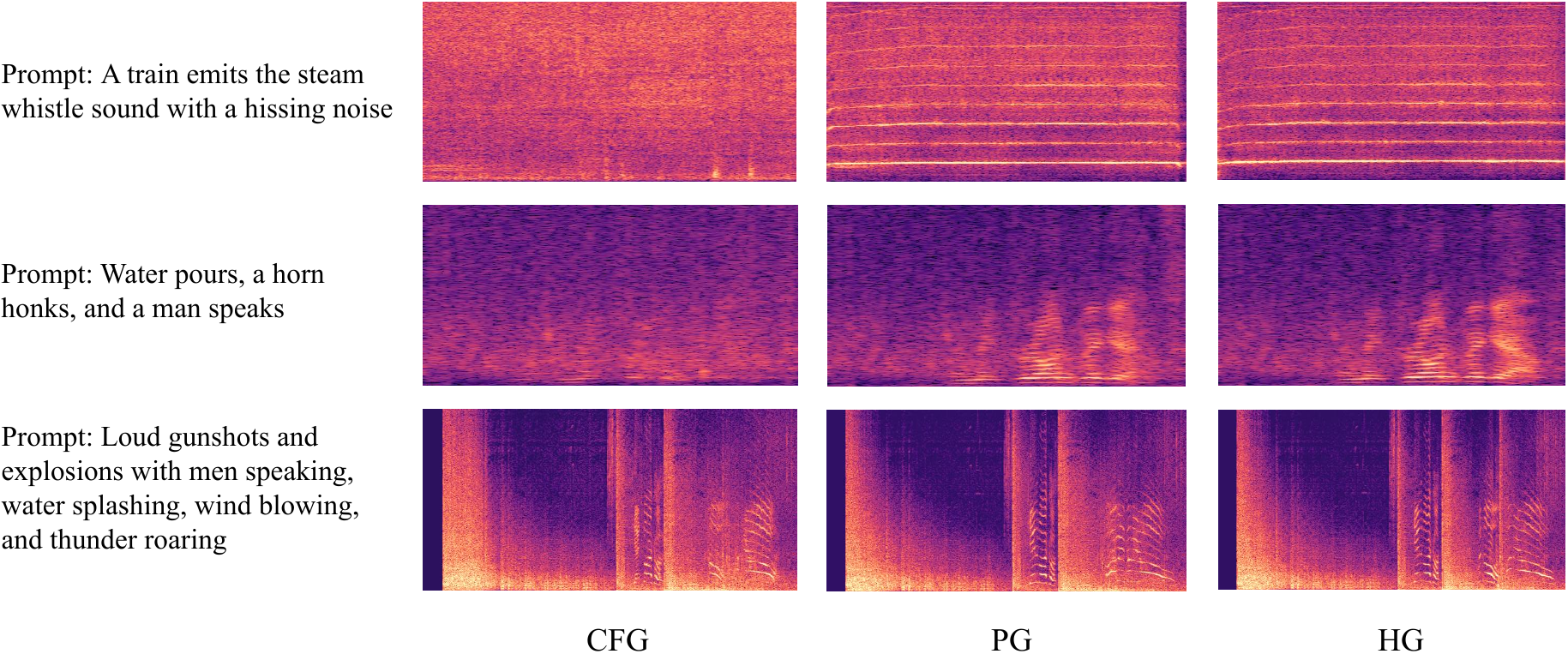}
  \caption{More T2A results comparing the spectrograms of the generated samples with different guidance strategies (CFG, PG, and HG) under various text prompts. The third sample is shown with a different time interval than the one presented in the main paper, and they share the same text prompt.}
  \label{fig:morecase}
\end{figure*}

\begin{figure*}[htbp]
  \centering
  \begin{subfigure}{0.48\textwidth}
    \centering
    \includegraphics[width=\linewidth]{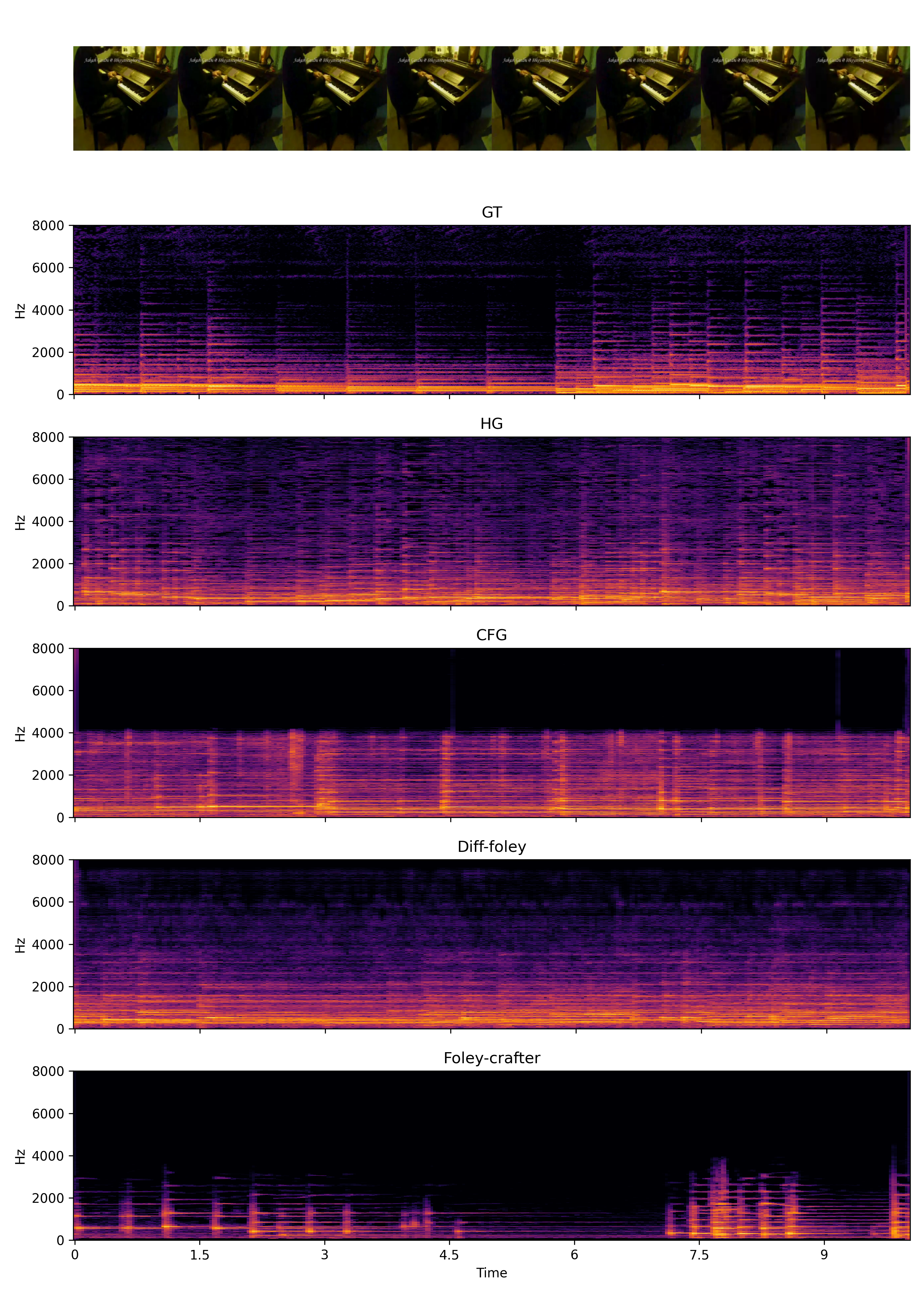}
    \caption{HG demonstrates the best generation quality.}
    \label{fig:left}
  \end{subfigure}
  \hfill
  \begin{subfigure}{0.48\textwidth}
    \centering
    \includegraphics[width=\linewidth]{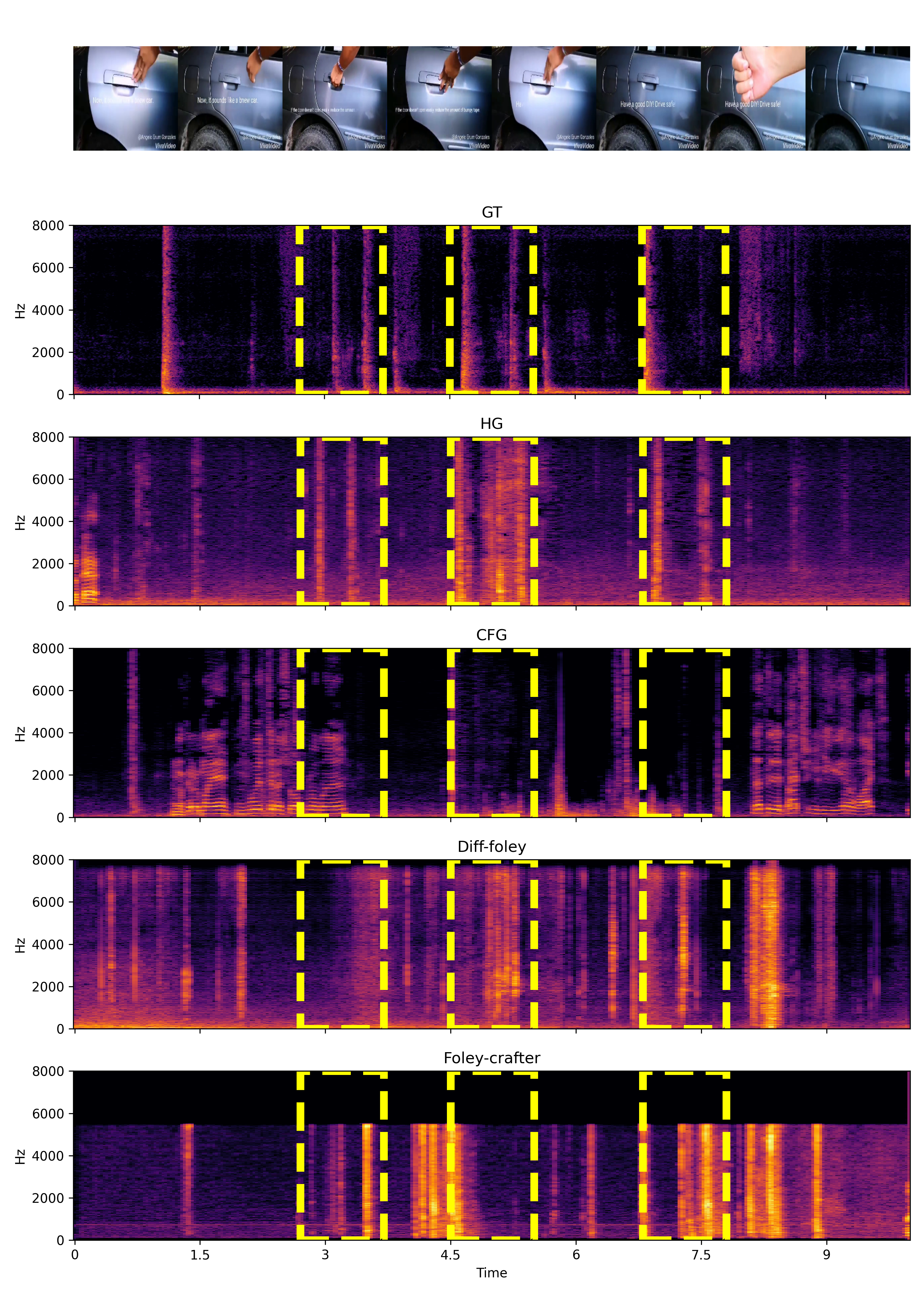}
    \caption{HG produces the most temporally aligned results.}
    \label{fig:right}
  \end{subfigure}
  \caption{More V2A comparing the spectrograms of the generated samples with different guidance strategies (CFG and HG) and baselines (Diff-foley and Foley-crafter).}
  \label{fig:v2amorecase}
  \vspace{-0.05in}
\end{figure*}

We shown more text-to-audio and video-to-audio generation results in Fig. \ref{fig:morecase} and \ref{fig:v2amorecase}, respectively. As shown in Fig. \ref{fig:morecase}, HG and PG consistently achieve more accurate harmonic modeling and superior temporal alignment compared to CFG in the first example. Their outputs exhibit well-defined harmonic stacks and consistent overtone structures, even in complex or polyphonic cases in the third example. Moreover, HG and PG maintain precise timing across events, effectively capturing the onset and duration of audio elements in the second example. In contrast, CFG often fails to organize harmonics coherently and produces temporally smeared results. These comparisons clearly illustrate the advantage of our methods in reinforcing both spectral clarity and temporal fidelity. In Fig. \ref{fig:v2amorecase}, HG generates significantly clearer high-frequency content and overall higher-quality audio compared to CFG in the first example. The resulting audio exhibits more natural brilliance and detail in the upper frequency range, enhancing perceptual realism. For the second example, HG demonstrates superior temporal alignment, accurately synchronizing audio events with visual cues, while CFG shows noticeable temporal drift and inconsistent timing. These comparisons further highlight the strengths of our method in improving both spectral resolution and temporal coherence in V2A generation. For more generated samples, please refer to our demo page: \url{audiomog.github.io}.

\section{Broader societal impact}
\label{societal}
Generative audio modeling presents substantial potential for misuse, which could result in harmful societal consequences. 
Principal concerns involve the dissemination of disinformation and the reinforcement of stereotypes and existing biases. 
Although our improvements enhance the realism and quality of generated samples, thereby potentially making misuse more convincing, they do not introduce any new capabilities or applications beyond those that already exist.

\section{Licenses}
\label{licenses}
\begin{itemize}[leftmargin=10pt]
    \item EDM2 models~\cite{karras2024analyzing, karras2024guiding}: Creative Commons BY-NC-SA 4.0 license
    \item Stable Audio Tools \cite{evans2024fast}: MIT license
    \item AudioLDM-Eval \cite{liu2023audioldm}: MIT license
    \item Diff-foley models \cite{luo2023diff}: Apache-2.0 license
    \item VTA-LDM models \cite{xu2024vtaldm}: Apache-2.0 license
    \item FoleyCrafter models \cite{zhang2024foleycrafter}: Apache-2.0 license
    \item V2A-Mapper models \cite{wang2024v2amapper}: Creative Commons BY-NC-ND 4.0 license
\end{itemize}

\end{document}